\documentclass[12pt]{article}
\usepackage{geometry}
\usepackage{a4}
\usepackage{graphicx}
\usepackage{epsf}
\usepackage{amsmath}
\usepackage{amssymb}
\usepackage{cite}
\newcommand{\be}{\begin{equation}}
\newcommand{\ee}{\end{equation}}

\newcommand{\Rmnum}[1]{\expandafter\@slowromancap\romannumeral #1@}
\newcommand{\bea}{\begin{eqnarray}}
\newcommand{\eea}{\end{eqnarray}}
\begin{document}
\def\C{{\mathbb{C}}}
\def\R{{\mathbb{R}}}
\def\s{{\mathbb{S}}}
\def\T{{\mathbb{T}}}
\def\Z{{\mathbb{Z}}}
\def\W{{\mathbb{W}}}
\def\Bbb{\mathbb}
\def\BZ{\Bbb Z} \def\BR{\Bbb R}
\def\BW{\Bbb W}
\def\BM{\Bbb M}
\def\BC{\Bbb C} \def\BP{\Bbb P}
\def\CP{\BC\BP}
\begin{titlepage}
\title{On the Higgsing and UnHiggsing of Fano 3-Folds}
\author{}
\date{
Prabwal Phukon, Tapobrata Sarkar
\thanks{\noindent E-mail:~ prabwal, tapo @iitk.ac.in}
\vskip0.4cm
{\sl Department of Physics, \\
Indian Institute of Technology,\\
Kanpur 208016, \\
India}}
\maketitle
\abstract{
\noindent
We study the Higgsing and UnHiggsing of M2-brane theories that probe cones over smooth toric Fano 3-folds, via brane tilings. We find many new examples of M2-brane 
gauge theories not studied previously in the literature, including those that do not correspond to cones over Fano varieties. 
Our analysis also provides, upon unHiggsing the known Fano varieties, new examples of quiver gauge theories that describe the same toric variety, but
with external point multiplicities. As a byproduct of our results, we discuss an example of a CY 4-fold that does not have a tiling description, and study the 
Higgsing of this theory.}
\end{titlepage}

\section{Introduction}

In recent years, there has been a lot of interest in understanding supersymmetric $(2+1)$d quiver Chern-Simons (CS) theories that correspond to world-volume theories on
M2-branes on various backgrounds, in the context of the ${\rm AdS}_4/{\rm CFT}_3$ correspondence. The key idea originated from the work of 
Schwarz \cite{schwarz}, who showed that higher supersymmetries than the well known $N=3$ example in $(2+1)$d theories can be achieved  by turning on CS couplings in the
field theory and setting the gauge kinetic terms to zero in a suitable limit. This resulted in a flurry of activity which culminated in the work of 
Aharony, Bergman, Jafferis and Maldacena \cite{abjm}, 
who constructed an $N=6$ supersymmetric quiver CS theory, following the work of Bagger and Lambert \cite{baglam}, and Gustavsson \cite{gusta}, 
which was conjectured to be the low energy theory on a stack of M2-branes probing a $\BZ_k$ orbifold of $\BC^4$. Since then, several authors have explored situations
with different supersymmetry, which are conjectured to be dual to M2-branes probing other backgrounds (see, e.g \cite{hoso1},\cite{hoso2},\cite{schnabl},\cite{baglam2}). 

In particular, a lot of focus has been on the construction of $(2+1)$d theories with $N=2$ supersymmetry, initiated in the works of \cite{ms},\cite{hanzaf},\cite{uedayama}, 
which are natural cousins of the well studied $N=1$ theories of D3-brane world volumes, that appear in the context of the ${\rm AdS}_5/{\rm CFT}_4$ duality. 
Initially, it was shown that these may arise naturally from ``similar'' D3-brane theories in $(3+1)$d,
i.e, having the same quiver diagram and superpotential, with the additional feature being the CS levels (which play the role of coupling constants) associated with the gauge groups. 
This was then extended to more generic situations, i.e for cases where the $(2+1)$d theory does not have a $(3+1)$d parent \cite{franco}.  As a result of these studies, 
it has been possible to understand a whole class of $N=2$ quiver CS theories in $(2+1)$d, that are conjectured to be theories on M2-branes transverse to Calabi-Yau (CY) 4-folds. 

In a recent paper \cite{fanohan}, Hanany and collaborators have constructed M2-brane theories which probe a class of toric CY 4-folds, which are 
complex cones over smooth Fano 3-folds. Fano
varieties are, by definition, algebraic varieties with the condition that the anti-canonical sheaf is ample. The classification of smooth Fano varieties has been known for sometime in
the mathematics literature. In two complex dimensions, there are five such varieties that are toric. In three complex dimensions, the situation is far more intricate, and it has been
shown that there are eighteen smooth Fano varieties (see eg. \cite{ww},\cite{baty1}). The M2-brane theories constructed in \cite{fanohan} 
(it was possible to construct 
fourteen of the possible eighteen) describe complex cones over such varieties, which can consequently be described as CY 4-folds that admit a brane tiling 
description (for a review, see \cite{kennaway}). \footnote{For the nomenclature of the smooth Fano varieties, we will use the conventions of \cite{fanohan} to 
which the reader is referred to for the quiver diagrams and brane tiling pictures.}

Naturally related to, and important in the understanding of M2-brane theories probing cones over Fano varieties, 
are  issues of Higgsing or unHiggsing of the same. These techniques often help in uncovering new gauge theories, and can further give rise to important ideas 
like dualities in gauge theories, the most famous example being Seiberg duality in the context of D3-branes. 
Simply speaking, in the Higgsing process, certain bifundamental fields are given vacuum
expectation values (vev s), so that one classically integrates out the fields that become massive as a result, and studies the moduli space of the resultant. 
Of course giving vev s to the fields introduces new Higgs scales in the theory, and the assumption here is that we flow to a scale that is much smaller compared to those
set by the vev s, in which limit the theory is again conformal. In the $(2+1)$d quiver CS cases, the resulting theories can often be
related to (phases of) other M2-brane theories or to new examples of CS theories \cite{hanhigg},\cite{bhs}. The reverse procedure,
called unHiggsing, aims to add fields to an existing theory with the constraint that the resulting theories satisfy the known properties of matter content and interactions as appropriate
for a $(2+1)$d CS theory \cite{bhs}. 

It is a natural question to ask whether one can construct sensible toric M2-brane theories by, say, unHiggsing those that probe Fano 3-folds. This is
related to the similar situation of $(3+1)$d theories, where it was possible to construct toric pseudo del Pezzo surfaces (which are $\BP^2$s blown up at more than
three non-generic points) \cite{unhigdp} as a result of unHiggsing the known del Pezzos. As we will show in the sequel, the answer to this is in the positive, and that 
the unHiggsing process for M2-brane theories describing cones over Fano 3-folds may result in interesting new gauge theories. In the process, 
we also obtain theories with different matter contents and superpotentials, that describe the same toric Fano 3-fold. Apart from being new phases for these theories, this is 
also interesting as it addresses some issues relating to the multiplicities of external points \footnote{We mean the multiplicity of {\it corner points} or {\it vertex points} in toric diagrams. 
Points in toric diagrams that are internal to a face or an edge i.e are not corner points might have multiplicities.} in toric diagrams.

The second issue that we address in this paper is that of Higgsing the Fano 3-folds. We show that for several examples, it is possible to Higgs a given Fano 3-fold theory
to a different one, but in many cases, Higgsing gives rise to a non-Fano varitey. In this study, we use the algorithm of \cite{tapo1} specialised to the case of CY 4-folds. Related
to this is the idea of the inverse algorithm that attempts to construct a sensible gauge theory living on an M2-brane given a toric CY 4-fold. In this paper, using the
inverse algorithm (which is generically almost impossible to tract, given its ambiguities, as we will elaborate in the text) we construct an example which we believe to be
an M2-brane theory that does not have a tiling description. We further consider Higgsing of this theory and show that it flows to a theory that admits brane tiling. 

The complete picture of Higgsing and unHiggsing Fano 3-folds, in order to map out the full set of theories that can be reached by these procedures, is admittedly a complicated
issue to deal with. We are hopeful that our work will initiate such a study. In particular, blowup and blowdown relations between these Fano varieties are known in the
mathematics literature, although all of them may not be tractable by Higgsing or unHiggsing. Nevertheless, as we will see, our analysis points to the fact that there possibly
exist different phases of some of these theories which are yet to be discovered. 

The paper is organised as follows. In section 2, we set up the notations and conventions to be used in the rest of the paper, and also provide some results on the 
unHiggsing of $(3+1)$d D3-brane theories that we believe have not appeared in the literature. In section 3, we perform Higgsing of Fano 3-fold theories and explicitly work out 
some examples, where the resulting theory may or may not be a Fano 3-fold. Section 4 deals with the unHiggsing of M2-brane theories and we present several new gauge 
theories which result from our analysis. Section 5 deals with a CY 4-fold theory that arises from the toric inverse algorithm, which does not seem to have a brane tiling picture. 
We also study some examples of Higgsing this theory. Finally section 6 concludes this paper with a summary of our results and future directions. 

 A word of caution before we proceed. In one of the original classification of Fano 3-folds \cite{ww}, the series of blow ups and blow down relations that exist between these
have been tabulated. Whereas using the results of \cite{fanohan} we recover a subset of them, we are not able to reproduce the full list, inspite of our efforts to 
find different phases of the theories constructed in \cite{fanohan}, although we believe that these should exist. Further, in several examples, the generators of the toric cones
of \cite{ww} do not seem to be related by a simple $GL(4,\BZ)$ transformation with those of \cite{fanohan}. We will proceed keeping these in mind, although to us this
indicates a necessity of more detailed study of M2-branes probing cones over Fano 3-folds.

\section{Higgsing and UnHiggsing of Fano 2 Folds}

In this section, we recapitulate some known facts about the Higgsing and UnHiggsing of Fano 2-folds probed by D3-branes transverse to the cone over the Fano variety. 
This section is mainly intended to serve as a review section, where we introduce the notations and conventions to be followed in the rest of the paper. However,
this section also contains some new results on multiplicities in toric diagram for D3-brane world volume theories, which, to the best of our knowledge has not been reported before. 
Let us begin by a brief review of the ``forward algorithm,''  which was originally developed for D3-branes probing orbifolds, in the work of \cite{dm},\cite{dgm}. 

Simply put, the forward algorithm deals with constructing the toric data of a singularity, given the information about the 
gauge theory living on a D-brane, with the brane being transverse to the singularity. The gauge theory is characterised by its matter content,
captured by the D-terms in the theory, and the superpotential, described by F-terms. For a single brane transverse to the singularity, the gauge fields are
bifundamental in nature, and the matter content is described by a quiver diagram, from which one can read off the charge matrix for the theory. The F-terms
on the other hand are not all independent, and can be solved in terms of a reduced number of fields. Labeling the latter by $v_j,j=1\cdots r+1$ (where
$r$ is the rank of the orbifolding group for D3-branes transverse to orbifold singularities), we can introduce a matrix $K$ that encodes the F-term constraints
as
\begin{equation}
X_i = \prod_j v_j^{K_{ij}}
\label{kmatrix}
\end{equation}
where the index $i$ runs over all the fields in the theory. Having obtained $K$, one calculates a dual matrix $T$ with the constraint that ${\vec K}.{\vec T} \geq 0$,
in terms of which the fields $v_j$ are written as
\begin{equation}
v_j = \prod_a p_a^{T_{ja}}
\end{equation}
where $p_a, a=1\cdots c$ denotes a new set of fields defined from $T$, and are identified with the fields in a four dimensional Gauged Linear Sigma Model (GLSM). 
The index $c$ has to be calculated on a case by case basis. Having written the $r+2$ $v_j$s in terms of $c$ fields, we then introduce an extra 
$U(1)^{c-r-2}$ gauge group,  and gauge invariance of the $v_j$ determine that  the charges of the fields under the new gauge group are encoded in a matrix
$Q$ that satisfies $T. Q^t=0$.  It is also be shown that the charges of the GLSM fields under the original gauge group is given by a matrix $VU$, such that
\begin{equation}
V. K^t=\Delta,~~~U.T^t= I
\end{equation}
where $\Delta$ is the charge matrix obtained from the quiver diagram, with a redundant $U(1)$ corresponding to the c.m motion of the branes, removed. 
Concatenating $Q$ (also called $Q_F$ since it encodes the superpotential constraints) and $VU$(also called $Q_D$ as it originates from the D-terms of the gauge
theory), and taking the kernel of the resulting matrix, we obtain the toric data of the singularity probed by the D3-brane, which we will denote by ${\cal T}$ throughout
this paper. 

A modern, superior version of the algorithm \cite{taxonomy} is via the language of brane tilings \cite{hanken}, \cite{kennaway}. In this approach, given the 
superpotential of a gauge theory (which need not be a partial resolution of an orbifold singularity), 
one directly constructs the ``perfect matching matrix'' $P$ which encodes information about the perfect matchings in the dimer model description of the theory \cite{hanken}. 
The kernel of this matrix is the matrix $Q_F$ of the previous paragraph. Further, the matrix  $Q_D$ is constructed as a suitable combination of
$P$ and the quiver charge matrix $d$. Specifically, upon introduction of a $1 \times G$ row matrix $C$ with all elements being unity
(G is the number of gauge groups in the theory), it was shown \cite{taxonomy} that the matrix $Q_D = {\rm kernel}\left(C\right). {\tilde Q}$ where the matrix ${\tilde Q}$ is one that
satisfies $d={\tilde Q}. P^t$. An advantage of this method is that it is directly applicable to M2-brane world volume theories with minimal modification. 

Recall that for $(2+1)$d quiver CS theories, one begins with a Lagrangian (in $N=2$ superspace notation) \cite{ms},\cite{hanzaf}
\begin{eqnarray}
{\mathcal L} &=& \int d^4\theta {\rm Tr}\left[\sum_{X_{ab}} X^{\dagger}_{ab} {\rm e}^{-V_a} X_{ab} {\rm e}^{V_b} + \sum_a \frac{k_a}{2\pi}
\int_0^1 dtV_a {\bar D}^{\alpha}\left({\rm e}^{tV_a}D_{\alpha}{\rm e}^{-tV_a}\right) \right] \nonumber\\
&+& \int d^2\theta W\left(X_{ab}\right) + cc
\end{eqnarray}
Where $a$ denotes the gauge group label, $V_a$ are vector superfields, $D_{\alpha}$ the superspace derivatives, and the $X_{ab}$ are the chiral bifundamentals. 
We will be mostly concerned with Abelian gauge groups. The $k_a$ denote the CS levels, with the constraints that
\begin{equation}
\sum_a k_a = 0
\end{equation}
which is necessitated by the fact that we are interested in CY 4-folds. Also, the choice of levels is made such that the greatest common divisor of the $k_a$ s is unity. The classical
moduli space of the theory is determined from the ``usual'' F and D-term constraints as in $(3+1)$d case, which for Abelian theories, reads
\begin{equation}
\partial_{X_{ab}}W = 0,~~~ \mu_a = \sum_a d_a |X|^2 = k_a \sigma
\label{dfconstraints}
\end{equation}
where the scalar component of the vector superfield, $\sigma_a$ are all set to a common value $\sigma$. Here, $d_a$ denotes the quiver charge of the bifundamental fields.
One can now take $G-2$ linear combinations of the second equation in (\ref{dfconstraints}) to identify the baryonic symmetries, corresponding to a vanishing l.h.s in the same.

For such theories, which are believed to be worldvolume theories on M2-branes, the forward algorithm needs to be modified as follows \cite{taxonomy}. 
Forming the $2\times G$ matrix $C$, all 
the elements of whose first row equal unity and the second row is formed by the CS levels for the gauge groups, the toric data 
obtained by concatenating $Q_D$ and  $Q_F$ (obtained as described before) results in the description of a CY four-fold, transverse to the brane world volume.

Having obtained a CY 3-fold or 4-fold theory, the process of Higgsing or unHiggsing can be effected by giving vev s and subsequent masses to certain bifundamentl
fields or adding new massless fields to the theory. In the process, one
either reduces or increases the number of gauge groups of the original theory. The process of Higgsing can be described directly in terms of the brane tiling picture
as follows \cite{tapo1}. We can remove (i.e give a vev to) a bifundamental field (or equivalently, a set of toric data points) from a theory, from its perfect matching matrix $P$. 
Removing the bifundamental field effectively removes all the perfect matchings in which this field takes part, and hence we obtain a reduced matching matrix, 
whose kernel gives the reduced $Q_F$ matrix of the theory. The dual of the kernel of the reduced $Q_F$ matrix now gives the reduced
$K$ matrix of the theory, defined in eq.(\ref{kmatrix}). This can now be used to construct the superpotential, and equivalently, the perfect matching matrix. Further, the face 
symmetries (combinations of fields that go around faces of the tiling, which in turn can be represented as combinations of the perfect matchings) 
of the resulting theory gives us the quiver charge matrix. This information, put together, is enough to determine the toric data of the resulting theory. Note that in this
method, we remove points from a toric diagram without embedding it into some known theory as in \cite{fhh}. Let us illustrate 
this by an example of Higgsing a del Pezzo surface. This has been extensively studied in the string theory literature for over the past decade.

As is known, the del Pezzo surfaces are compact divisors in CY 3-folds which are thus described as cones over these surfaces. There are ten such surfaces of which 5 has been 
known to be toric. These are $\BP^1\times \BP^1$, $\BP^2$ and $\BP^2$ blown up at one, two and three generic points that are denoted by $dP_0$, $dP_1$ and $dP_2$ 
respectively. In \cite{unhigdp}, it was shown that it is possible to obtain more examples of toric del Pezzos by unHiggsing the surface $dP_3$, 
$\BP^2$ blownup at more than three non-generic points. These were named the pseudo del-Pezzo surfaces $PdP_4$ and $PdP_5$. 

From the nomenclature, it is clear that for example, the  cone over $dP_2$ is the resolution of the cone over $dP_3$. To illustrate the Higgsing procedure of \cite{tapo1}, we 
choose this example. The cone over the third del Pezzo surface, $dP_3$ is a theory of 6 gauge groups and 12 chiral multiplets.
The superpotential of the theory is 
\begin{eqnarray}
W &=& X_{12} X_{23} X_{34} X_{45} X_{56} X_{61} + X_{13} X_{35} X_{51}  + X_{24} X_{46} X_{62}\nonumber\\
&-& X_{23} X_{35} X_{56} X_{62}- X_{13} X_{34} X_{46} X_{61} -
X_{12} X_{24} X_{45} X_{51}
\label{dp3superpot}
\end{eqnarray}
From the superpotential, we can construct the perfect matching matrix and this is given by
\begin{equation}
P = 
\begin{pmatrix}
& p_{1} & p_{2} & p_{3} & p_{4} &p_{5} & p_{6}& p_{7} & p_{8} & p_{9}& p_{10}& p_{11} & p_{12}  \\ 
X_{12} & 0 & 0 & 1 & 0 & 0 & 0 & 0 & 0 & 1 & 0 & 0 & 0\\ 
X_{13} & 1 & 0 & 0 & 0 & 0 & 1 & 0 & 0 & 1 & 1 & 0 & 0\\ 
X_{23} & 1 & 0 & 0 & 0 & 1 & 0 & 0 & 0 & 0 & 0 & 0 & 0\\ 
X_{24} & 1& 1 & 0 & 0 & 0 & 1 & 0 & 1 & 0 & 0 & 0 & 0  \\ 
X_{34} & 0 & 1 & 0 & 0 & 0 & 0 & 0 &0 & 0 & 0 & 1 & 0 \\ 
X_{35} & 0 & 1 & 1 & 1 & 0 & 0 & 0 & 1 & 0 & 0 & 0 & 0 \\
X_{45} & 0 & 0 & 0 & 1 & 0 & 0 & 0 & 0 & 0 & 1 & 0 & 0 \\ 
X_{46} & 0 & 0 & 1 & 1 & 1& 0 & 1 & 0 & 0 & 0 & 0 & 0  \\ 
X_{51} & 0 & 0 & 0 & 0 & 1 & 0 & 1 & 0 & 0 & 0 & 1 &1\\ 
X_{56} & 0 & 0 & 0 & 0 & 0 & 1 & 1 & 0 & 0 & 0 & 0 & 0 \\ 
X_{61} & 0 & 0 & 0 & 0 & 0 & 0 & 0 & 1 & 0 & 0 & 0 & 1 \\ 
X_{62} & 0 & 0 & 0 & 0 & 0 & 0 & 0 & 0 & 1 & 1 &1 & 1
\end{pmatrix}
\label{dp3}
\end{equation}
Now suppose we Higgs the theory by giving a vev to the field $X_{56}$. From our earlier discussion, the reduced matching matrix is obtained by removing the 10th row, 
and the 6th and 7th  columns from the matrix of eq.(\ref{dp3}). The dual of the kernel of the reduced matching matrix is then seen to be given by
\begin{equation}
K = 
\begin{pmatrix}
0 & 0 & 0 & 0 & 0 & 0  & 0 & 0 & 1 & 1 & 1 \\ 
0 & 0 & 0 & 0 & 0 & 0 & 0 & 1 & 0 & 1 & 1  \\ 
0 & 0 & 0 & 0 & 0 & 1 & 1 & 0 & 0 & 0 & 1 \\ 
0 & 0 & 0 & 0 & 1 & 0 & 1 & 0 & 0 & 0 & 1 \\  
0 & 0 & 1 & 1 & 0 & 0 & 0 & 0 & 1 & 0 & 0 \\ 
1 & 1 & 0 & 0 & 0 & 0 & 0 & 0 & 0 & 1 & 0  \\
0 & 1 & 0 & 1 & 0 & 1 & 0 & 0 & 0 & 0 & 0
\end{pmatrix}
\end{equation}
Integrating the $K$ matrix, one obtains the superpotential
\begin{eqnarray}
W &=& X_{10} X_{3} X_{6} X_{5} + X_{9} X_{2} X_{7} X_{8} + X_{11} X_{1} X_{4}  \nonumber\\
&-& X_{11} X_{2} X_{3} -X_{1} X_{5} X_{6} X_{8} X_{9} - X_{10} X_{4} X_{7}
\end{eqnarray}
The tiling is easily constructed (whence we can identify the quiver charges of the fields $X_{1\cdots 11}$) and one can check that there are five face symmetries 
(corresponding to five gauge groups) generated by
\begin{equation}
F_1 = p_8 - p_1,~F_2 = p_9 - p_8,~F_3 = p_1 - p_6,~F_4=p_6 - p_3,~F_5=p_3 - p_9
\end{equation}
Thus, one obtains the matrix
\begin{equation}
Q_D = 
\begin{pmatrix}
-1  & 0 & 0 & 0 & 0 & 0 & 0 & 1 & 0 & 0  \\ 
0 & 0 & 0 & 0 & 0 & 0 & 0 & -1 & 1 & 0 \\ 
1 & 0 & 0 & 0 & 0 & -1 & 0 & 0 & 0 & 0 \\  
0 & 0 & -1 & 0 & 0 & 1 & 0 & 0 & 0 & 0 
\end{pmatrix}
\end{equation}
Now noting that the kernel of the reduced perfect matching matrix gives the $Q_F$ matrix for the Higgsed theory, we obtain the toric data for the resulting theory as
\begin{equation}
{\cal T} = 
\begin{pmatrix}
0 & -1 & 0 & 0 & 1  & 0 & 0 & 0 & 0  & 1 \\   
0 & 0  & 0 & -1 & -1 & 0  & 1 & 0 & 0 & 0 
\end{pmatrix}
\label{toricdp2}
\end{equation}
The theory has 5 gauge groups and 11 bifundamental fields, and along with the toric data above is recognised to the the cone over the second del Pezzo surface, $dP_2$.
\begin{figure}[t!]
\begin{minipage}[b]{0.5\linewidth}
\centering
\includegraphics[width=3in,height=2.5in]{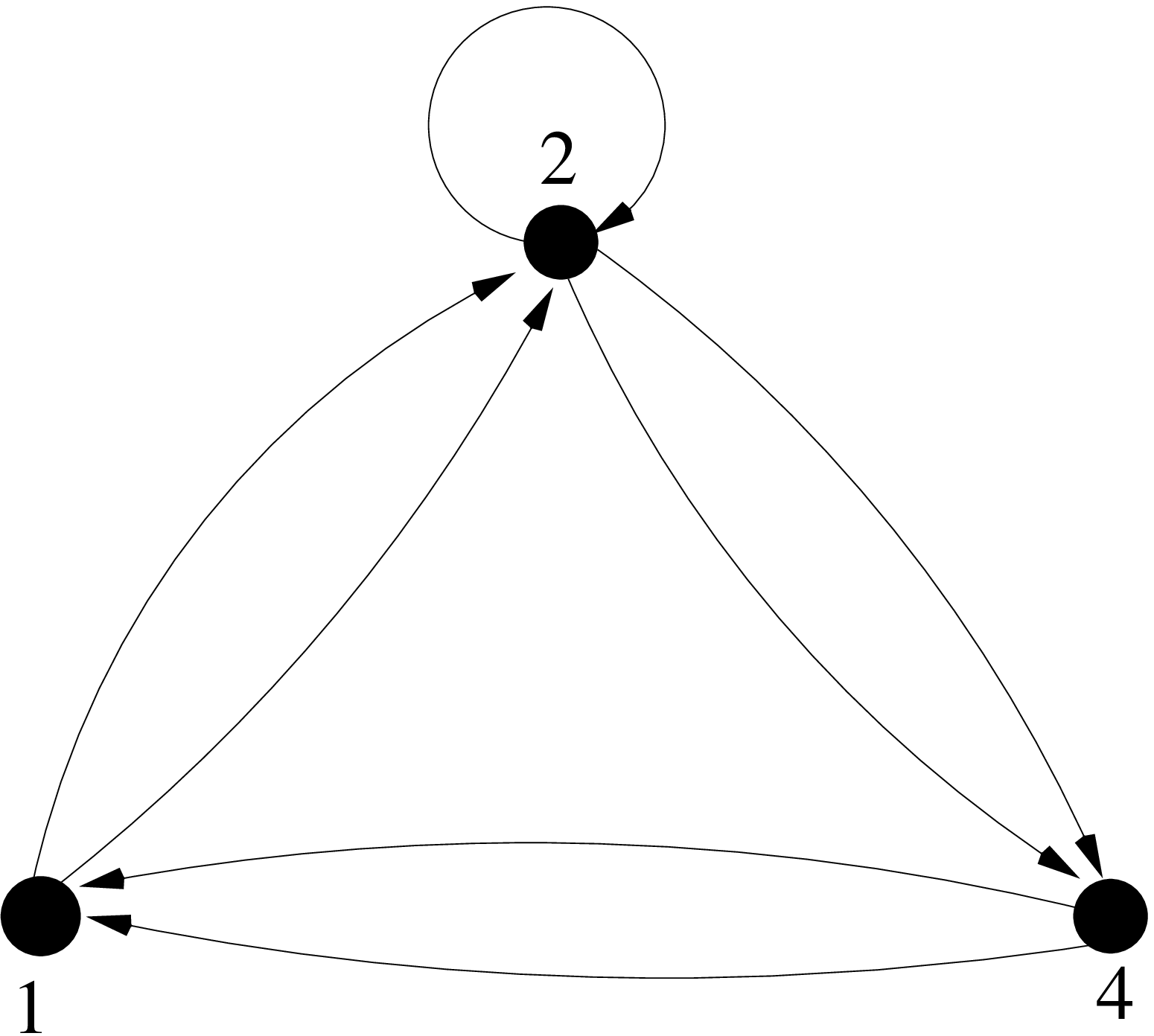}
\caption{UnHiggsing with $X_{23}$, $X_{34}$, $X_{42}$ gives the theory $dP_1$}
\label{unhigdp1}
\end{minipage}
\hspace{0.6cm}
\begin{minipage}[b]{0.4\linewidth}
\centering
\includegraphics[width=2in,height=2.5in]{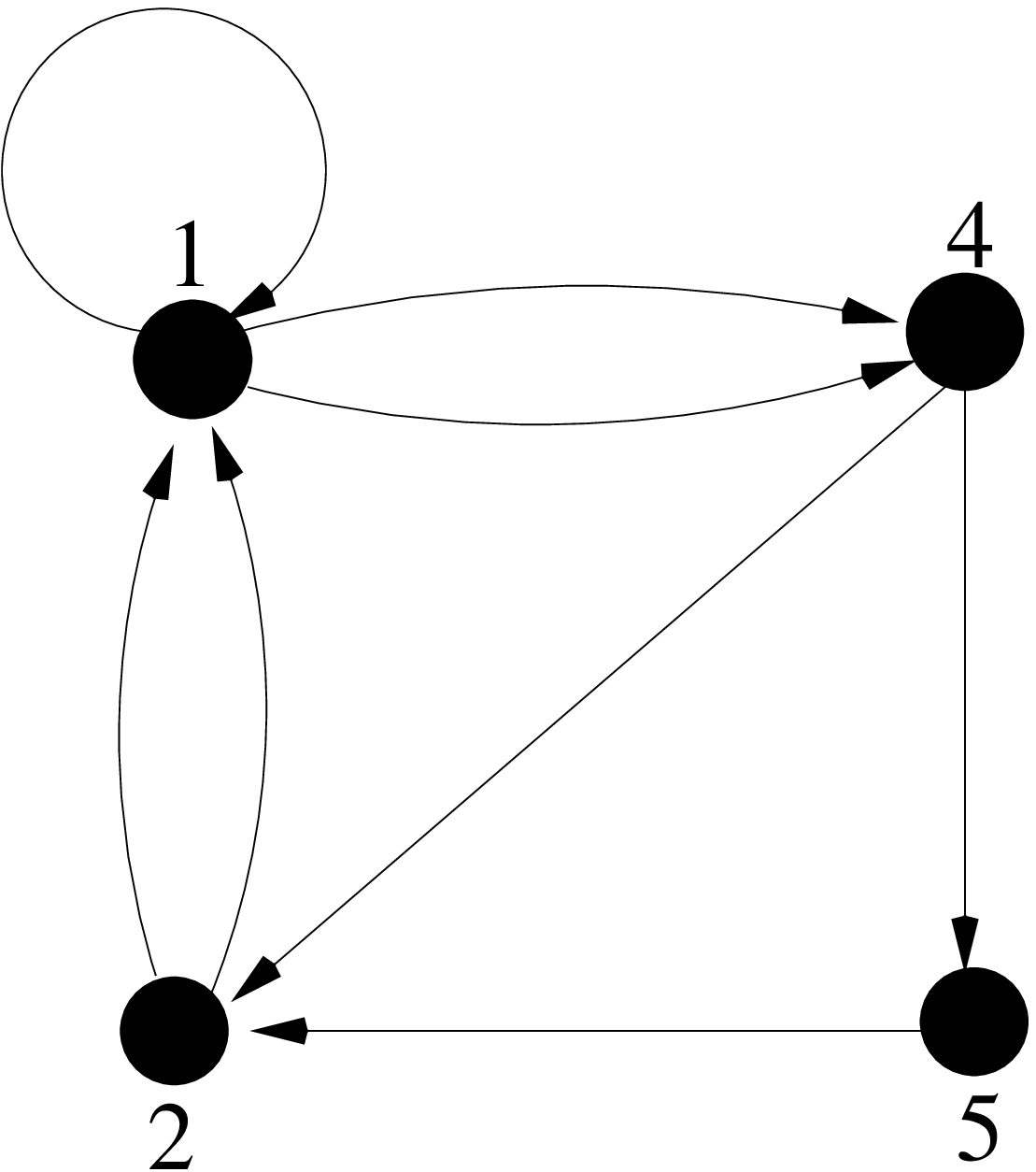}
\caption{UnHiggsing with $X_{31}$, $X_{15}$, $X_{53}$ gives the theory $dP_2$}
\label{unhigdp2}
\end{minipage}
\end{figure}
Let us pause to highlight the difference between the approach of \cite{tapo1} and that of, for example, \cite{hanhigg}, \cite{bhs}. In the latter, vev s are given to various 
combinations of fields in the original theory to reach the quiver and superpotential of a reduced theory, whence the forward algorithm is applied to obtain the vacuum 
moduli space of the latter. Our method, on the contrary, can be used to directly remove points from a toric diagram, and is thus closer in spirit to the inverse toric algorithm originally
introduced in the context of D3-brane theories in \cite{fhh}. It also naturally incorporates adjoints in the presence of which the dual of the kernel of the reduced 
matching matrix is typically inconsistent \cite{tapo1}. We will come to this point in more details later. 
\begin{figure}[t!]
\centering
\includegraphics[width=3in,height=2.5in]{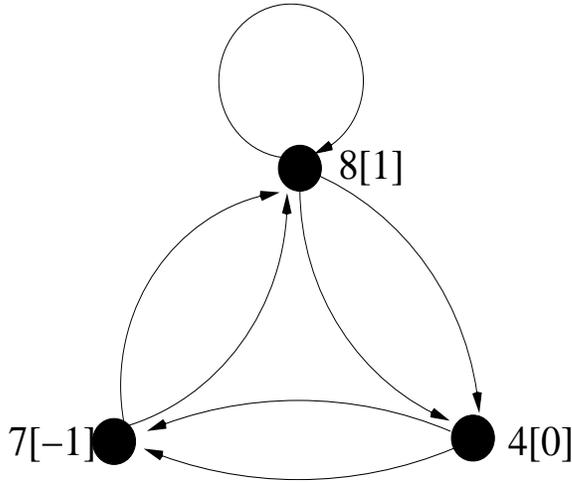}
\caption{Quiver diagram for toric data of eq.(\ref{oquivertoric}) }
\label{oquiver}
\end{figure}

As a further illustration, we study the Higgsing of the $(2+1)$d quiver CS theory, the phase II of the orbifold $\BC^4/\BZ_2^3$. Our notations follow \cite{bhs} where this 
has been worked out.The details of Higgsing quiver CS theories will be elaborated upon later, but let us just point out that using the method described above, 
we find that the theory with quiver diagram depicted in fig.(\ref{oquiver}) and having the superpotential 
\begin{equation}
W = X_{23}^1X_{31}^1X_{12} - X_{23}^2X_{31}X_{12} + X_{23}^2X_{31}^2X_{12}\phi - X_{23}^1X_{31}^1X_{12}\phi
\end{equation}
describes the toric variety given by the data
\begin{equation}
{\cal T} = 
\begin{pmatrix}
0&0&0&1&0&0\\
1&0&1&0&0&0\\
-1&1&0&0&0&0
\end{pmatrix}
\label{oquivertoric}
\end{equation}
and can simply be seen to be another phase of the theory described in \cite{bhs} (theory ``o'' of Table 5 of that paper).  

UnHiggsing toric 3-folds or 4-folds, which will also form a part of this paper, have also been well studied in the literature, beginning with the work of
\cite{unhigdp}. For M2-brane theories, this has recently been addressed in \cite{bhs}. In this approach, instead of removing fields from the parent theory, we add bifundamental
fields to the same, and modify the superpotential appropriately. The constraint here is that Higgsing of the theory with the added fields should reduce to the
theory one begins with. The procedure sometimes lead to interesting gauge theories living on D3 or M2-brane world volumes which otherwise may not be
obtained from partial resolutions, as we show in sequel. 

\begin{figure}[t!]
\begin{minipage}[b]{0.5\linewidth}
\centering
\includegraphics[width=3in,height=2.5in]{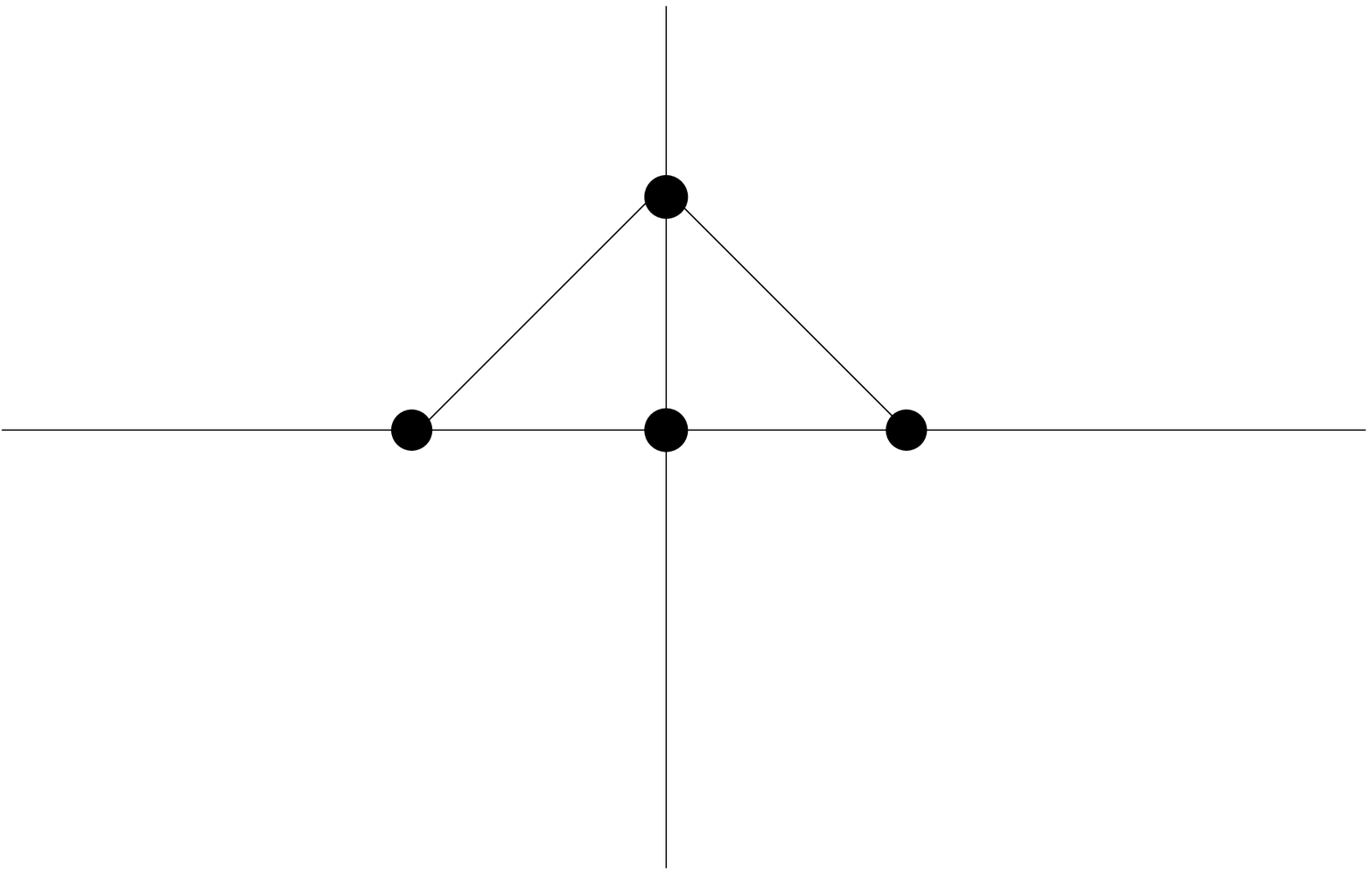}
\caption{Toric data for the theories mentioned in figs.(\ref{unhigdp1}) and (\ref{unhigdp2}).}
\label{commontoric}
\end{minipage}
\hspace{0.6cm}
\begin{minipage}[b]{0.5\linewidth}
\centering
\includegraphics[width=3in,height=2.5in]{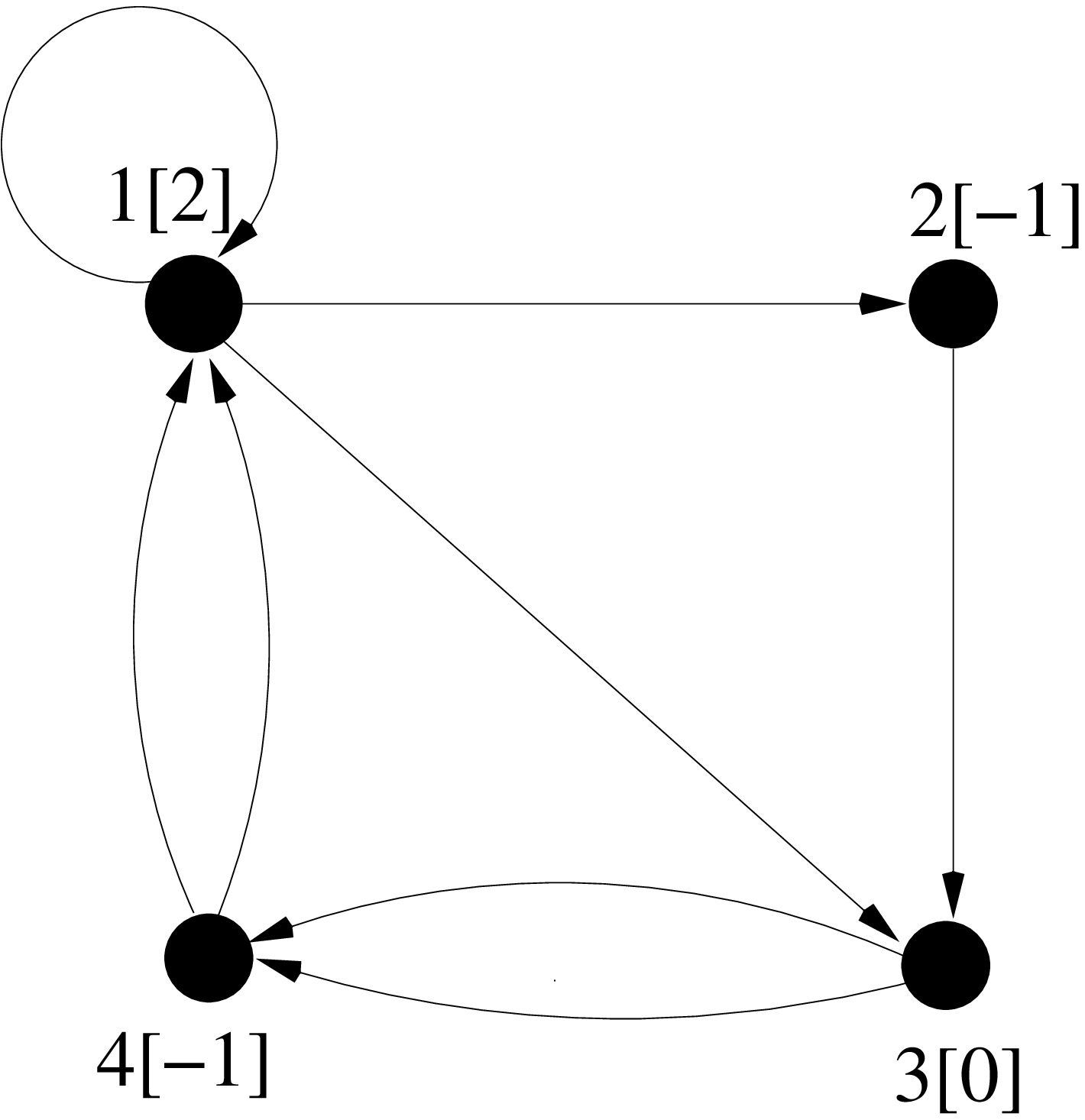}
\caption{UnHiggsing with $X_{51}$, $X_{14}$, $X_{45}$ gives the $(2+1)$d theory ${\mathcal E_1}$}
\label{unhige1}
\end{minipage}
\end{figure}

In fig.(\ref{unhigdp1}) and fig.(\ref{unhigdp2}), we have provided two examples of quiver gauge theories, which, on unHiggsing by 
three fields, give rise to the complex cones over the first and the second del Pezzo surfaces. For the quiver of fig.(\ref{unhigdp1}), the superpotential is taken to
be
\begin{equation}
W = X_{12}^1 X_{24}^2 X_{41}^1-X_{22} X_{12}^1 X_{24}^2 X_{41}^2- X_{12}^2 X_{24}^1 X_{41}^1+X_{22} X_{24}^1 X_{41}^2
X_{12}^2
\label{sup1}
\end{equation}
whereas for the quiver of fig.(\ref{unhigdp2}), the superpotential is 
\begin{equation}
W = X_{14}^1 X_{45} X_{52} X_{21}^2 X_{11}-X_{11} X_{14}^2 X_{45} X_{52} X_{21}^1-X_{42} X_{21}^2 X_{14}^1+ X_{42} X_{21}^1
X_{14}^2
\label{sup2}
\end{equation}
The first theory, on unHiggsing by the bifundamental fields $X_{23}$, $X_{34}$ and $X_{42}$ can be shown to give rise to the quiver gauge theory for the complex cone over
the first del Pezzo surface, while the second, on unHiggsing by the fields  $X_{31}$, $X_{15}$ and $X_{53}$ leads to the cone over the second del Pezzo surface. 
For future reference, we have given, in fig.(\ref{unhige1}), the quiver diagram of a theory which when unHiggsed by the fields $X_{51}$, $X_{14}$ and
$X_{45}$ gives us the theory on the toric fano 3-fold, ${\mathcal E_1}$. The CS levels for the gauge groups is given in the square brackets in the figure. 
For completeness, we mention that the toric data for the quivers of fig.(\ref{unhigdp1}) and (\ref{unhigdp2}) (in conjunction with the superpotentials given in eq.(\ref{sup1})
and (\ref{sup2}) are the same, and the toric diagram is presented in fig.(\ref{commontoric}).

\begin{figure}[t!]
\begin{minipage}[b]{0.5\linewidth}
\centering
\includegraphics[width=3in,height=2.5in]{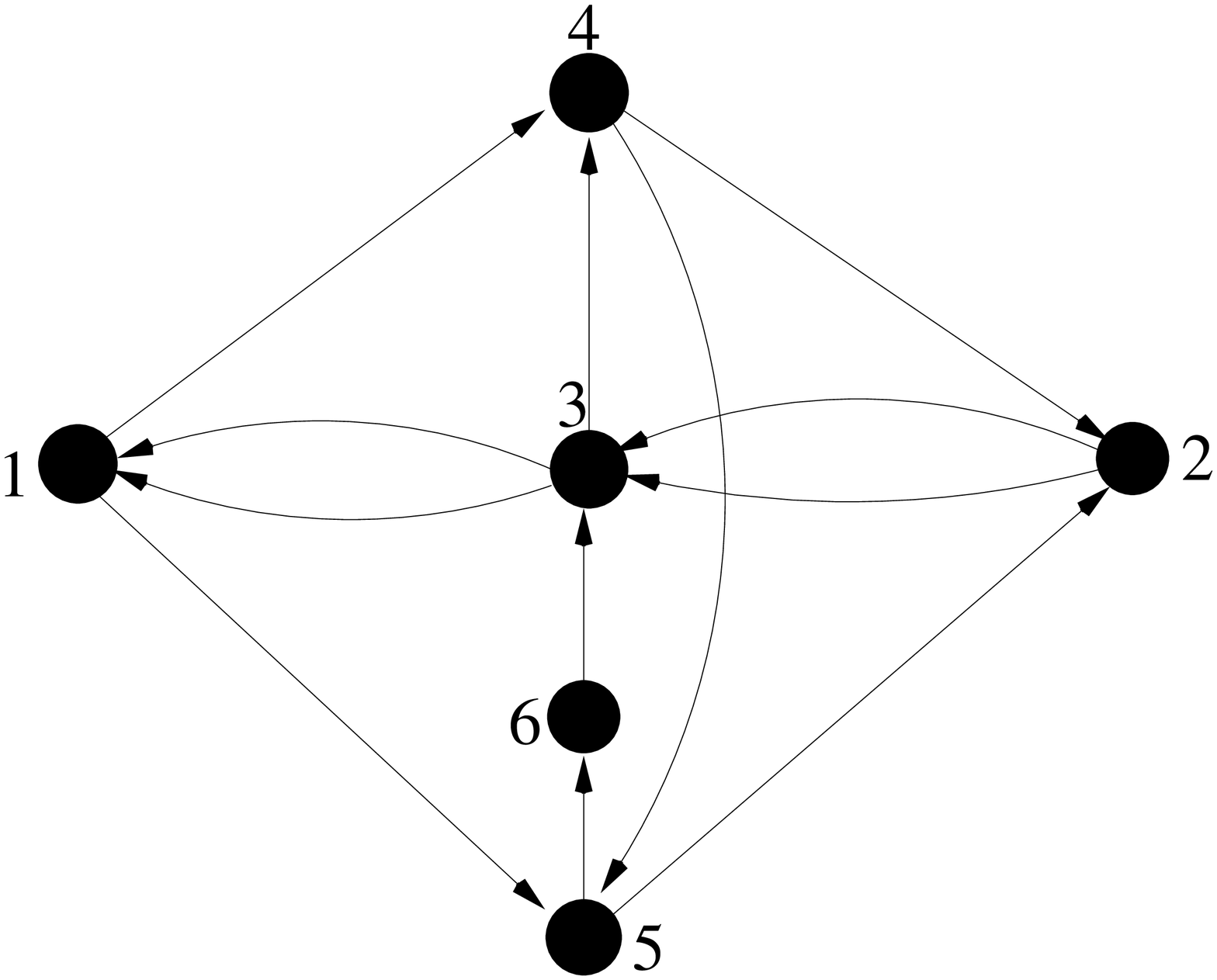}
\caption{Quiver diagram for unHiggsing $dP_2$ by the field $X_{56}$}
\label{qunhigdp2a}
\end{minipage}
\hspace{0.6cm}
\begin{minipage}[b]{0.5\linewidth}
\centering
\includegraphics[width=3in,height=2.5in]{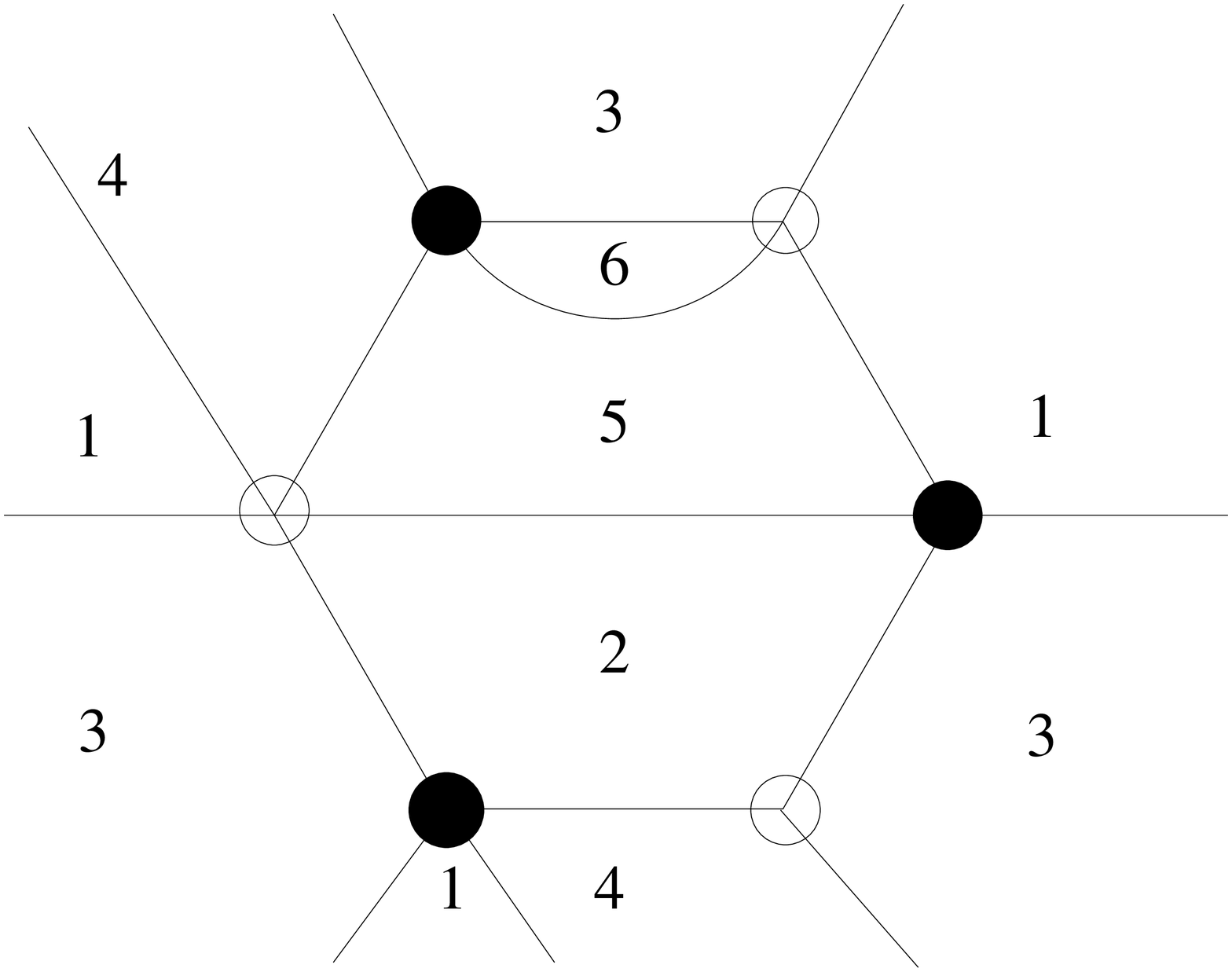}
\caption{Brane tiling for UnHiggsing $dP_2$ by the field $X_{56}$}
\label{tunhigdp2a}
\end{minipage}
\end{figure}

Before ending this section, let us point out that unHiggsing the del Pezzo theories might sometime lead to different D3-brane theories that represent the {\it same}
toric variety. These are interesting, as they do not seem to appear from partial resolutions of the orbifold $\BC^3/\BZ_3 \times \BZ_3$. Let us see if we can substantiate this.
As an example, consider unHiggsing the cone over $dP_2$ by the field $X_{56}$. The resulting superpotential reads
\begin{eqnarray}
W &=& X_{34}X_{45}X_{56}X_{63} + X_{15}X_{52}Y_{23}X_{31} + X_{14}X_{42}X_{23}Y_{31} \nonumber\\
&-& X_{14}X_{45}X_{52}X_{23}X_{31} - X_{34}X_{42}Y_{23}- X_{15}X_{56}X_{63}Y_{31}
\end{eqnarray}
The quiver and the tiling are presented in fig.(\ref{qunhigdp2a}) and (\ref{tunhigdp2a}). The toric data for this singularity can be seen to be the same as that of 
the complex cone over $dP_2$ given in eq.(\ref{toricdp2}) but now with different multiplicities, and is given by
 \begin{equation}
{\cal T} = 
\begin{pmatrix}
0 & -1 & 0 & 1 & 0 & 1 \\   
0 & 0  &  -1 & -1 & 1  & 0 \\
7 & 1 & 1& 1& 2& 2 
\end{pmatrix}
\label{toricdp2b}
\end{equation}
where the last row denotes the multiplicities of the toric data points. Note that two of the vertex point has multiplicity greater than one. 

\begin{figure}[t!]
\begin{minipage}[b]{0.5\linewidth}
\centering
\includegraphics[width=3in,height=2.5in]{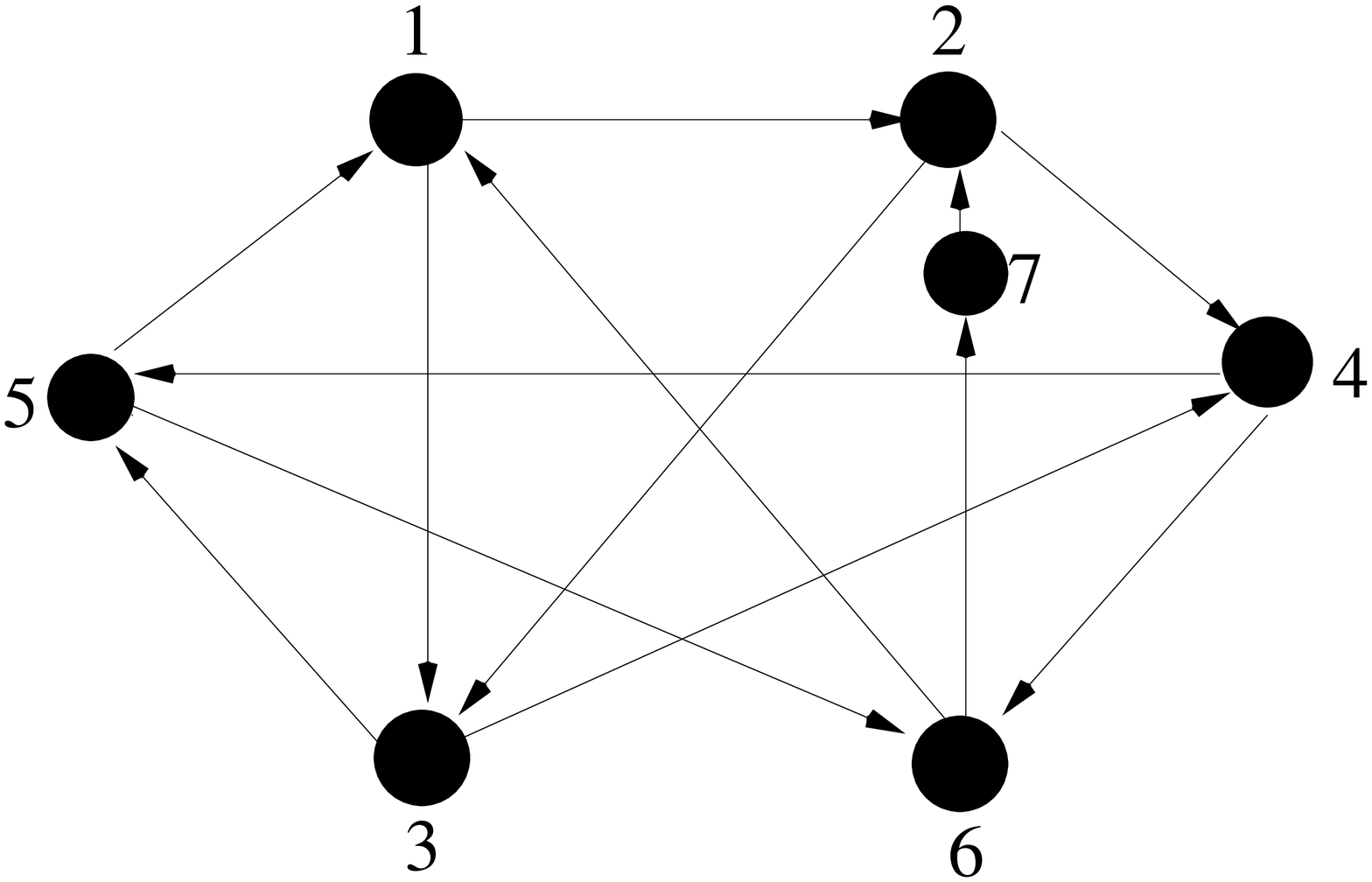}
\caption{Quiver diagram for unHiggsing $dP_3$ by the field $X_{67}$}
\label{qunhigdp3a}
\end{minipage}
\hspace{0.6cm}
\begin{minipage}[b]{0.5\linewidth}
\centering
\includegraphics[width=3in,height=2.5in]{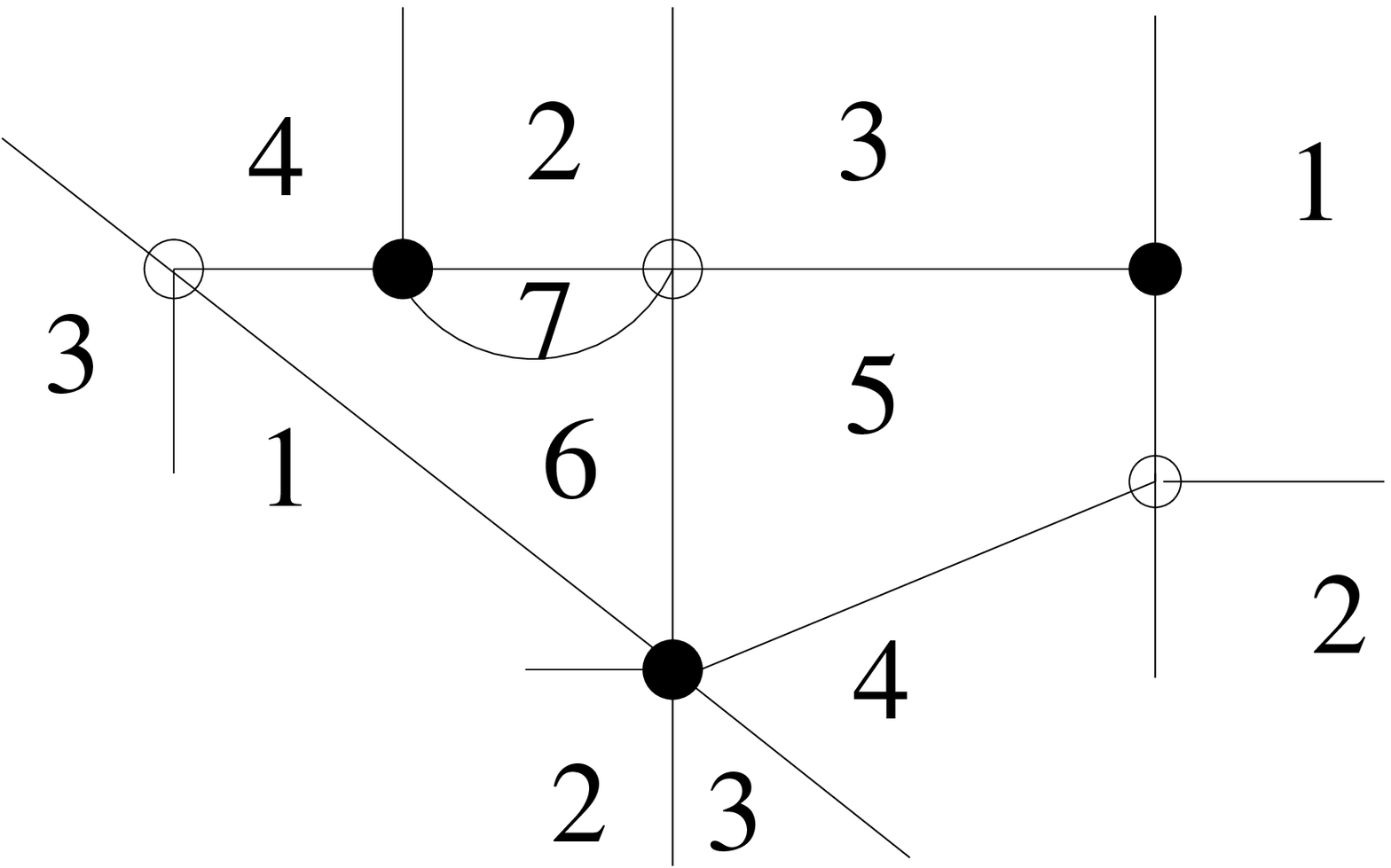}
\caption{Brane tiling for UnHiggsing $dP_3$ by the field $X_{67}$}
\label{tunhigdp3a}
\end{minipage}
\end{figure}

The second example is the unHiggsing of the cone over the $dP_3$ surface by the field $X_{67}$, in eq.(\ref{dp3superpot}). The superpotential is now modified to
\begin{eqnarray}
W &=& X_{12} X_{23} X_{34} X_{45} X_{56} X_{61} + X_{13} X_{35} X_{51}  + X_{24} X_{46} X_{67}X_{72}\nonumber\\
&-& X_{23} X_{35} X_{56} X_{67}X_{72}- X_{13} X_{34} X_{46} X_{61} -
X_{12} X_{24} X_{45} X_{51}
\label{dp3superpota}
\end{eqnarray}
The quiver data and the tiling for this model is given in fig.(\ref{qunhigdp3a}) and (\ref{tunhigdp3a}) respectively. The toric data again shows multiplicities in the
external points, and is given by
\begin{equation}
{\cal T} = 
\begin{pmatrix}
0 & 0& 1&-1&-1&0&1 \\   
0&-1&-1&1&0&1&0 \\
8&1&1&1&1&2&2
\end{pmatrix}
\label{toricdp3b}
\end{equation}
where again the last row denotes the multiplicities of the fields. Note that this theory also has vertex point multiplicities. 
We will encounter similar cases in CY 4-fold theories as well. 

Before we end, let us summarise the results in this section. Apart from reviewing standard procedure on Higgsing and unHiggsing of D3- and M2-brane
theories, we have seen that unHiggsing Fano 2-folds sometimes results in the new theory reproducing the same toric variety, but with different vertex point (or corner
point) multiplicities.
Note that the example considered here (the $dP_2$ and $dP_3$ surfaces) are isolated singularities, for which such multiplicities are not known to arise in the description of these
as partial resolutions of the orbifold $\BC^3/\BZ_3\times\BZ_3$. We have presented the quiver diagram and the brane tiling for these theories. We have also presented 
our first example of unHiggsing an M2-brane theory to a Fano 3-fold, to which we will turn to in details later. These are the main results of this section. 

\section{Higgsing Fano 3-Folds}

We are now ready to undertake Higgsing operations for Fano 3-folds. To remind the reader, our notations will follow \cite{fanohan}, which we briefly recall.
As we have said before, there are 18 known smooth toric Fano 3-folds. The simplest of them is the projective space $\BP^3$. The varieties $\BP^2\times \BP^1$ and
$\BP^1\times \BP^1\times \BP^1$ are called ${\mathcal B}_4$ and ${\mathcal C}_3$ respectively. The latter is also the $Q^{1,1,1}/\BZ_2$ theory which is known
to possess two distinct toric phases. The varieties $\BP^1\times dP_1$, $\BP^1\times dP_2$ and
$\BP^1\times dP_3$ where $dP_{1,2,3}$ are the del Pezzo surfaces are named ${\mathcal C}_4$, ${\mathcal E}_3$ and ${\mathcal F}_1$ respectively. ${\mathcal B}_2$
is the projectivisation of the total line bundle $\BP\left({\mathcal O}_{\BP^2} \oplus {\mathcal O}_{\BP^2}(1)\right)$. The $\BP^1$ blowups of ${\mathcal B}_2$ and
${\mathcal B}_4$ are the Fano varieties ${\mathcal D}_1$ and ${\mathcal D}_2$. The varieties ${\mathcal E}_{1,2,4}$ are $dP_2$ bundles over the
base $\BP^1$,  ${\mathcal F}_2$ is the $dP_3$ bundle over $\BP^1$ and the remaining, ${\mathcal B}_{1,2}$, ${\mathcal C}_{1,2,5}$ are appropriate projectivisations
of line bundles such that they are of complex dimension three.  Out of the 18 Fano 3-folds mentioned above, utilising the symmetry of the varieties, 
14 were shown to have a tiling description in \cite{fanohan}, with the exception of $\BP^3$ and ${\mathcal B}_{1,2,3}$. 
It was also shown that the toric data encodes the simple roots of the (non-Abelian) mesonic moduli space symmetry. 
The cones over the Fano 3-folds are the noncompact CY 4-fold conjectured to be probed by the M2-brane theory.
A necessary condition of this is that the toric data of the variety includes the origin. There are various blowup and blowdown relations between the
smooth Fano 3-folds \cite{ww} via contracting divisors to a line or a point and should be describable in string theory. 

As a warmup, and an illustration of the method described in the previous section, let us consider Higgsing the theory ${\mathcal D}_1$. We will keep in mind
that under a Higgsing process, the CS levels of the two gauge groups which collapse to one as a result of the Higgsing are added \cite{bhs}. We start with the 
perfect matching matrix of the ${\mathcal D}_1$ theory \cite{fanohan}
\begin{equation}
P = 
\begin{pmatrix}
& p_{1} & p_{2} & p_{3} & p_{4} &p_{5} & p_{6}& p_{7} & p_{8}\\
X_{41}^1 & 1&0&0&0&1&0&0&0\\
X_{41}^2&0&1&0&0&1&0&0&0\\
X_{23}^1&1&0&0&0&0&1&0&0\\
X_{23}^2&0&1&0&0&0&1&0&0\\
X_{34}^1&1&0&1&0&0&0&1&0\\
X_{34}^2&0&1&1&0&0&0&1&0\\
X_{34}^3&0&0&0&1&0&0&1&0\\
X_{12}&0&0&1&0&0&0&0&1\\
X_{42}&0&0&0&1&1&0&0&1\\
X_{13}&0&0&0&1&0&1&0&1
\end{pmatrix}
\end{equation}
Here, the CS levels are taken to be ${\vec k} = \left(-1,-1,0,2\right)$. 
If we give a vev to the field $X_{23}^2$, we remove the perfect matchings $p_2$ and $p_6$ and hence obtain a reduced matching matrix whose kernel directly
gives $Q_F = \left(0~1~1~0~-1~-1\right)$. We can calculate the new perfect matching matrix using the T, and its dual K matrices (described in section 2) now given by 
\begin{equation}
T =
\begin{pmatrix}
0&1&0&0&0&1\\
0&1&0&0&1&0\\
0&0&0&1&0&0\\
0&-1&1&0&0&0\\
1&0&0&0&0&0
\end{pmatrix},~
K =
\begin{pmatrix}
0&0&0&0&1&1&0\\
0&0&1&1&0&0&0\\
0&1&0&0&0&0&1\\
0&0&0&1&0&1&0\\
1&0&0&0&0&0&1
\end{pmatrix},~
P = 
\begin{pmatrix}
1&0&0&0&0&0\\
0&0&0&1&0&0\\
0&1&0&0&1&0\\
0&0&1&0&1&0\\
0&1&0&0&0&1\\
0&0&1&0&0&1\\
1&0&0&1&0&0
\end{pmatrix}
\label{Tmatrix}
\end{equation} 
Where, in the $K$ matrix, we have added a column to make it consistent, and the added field can be seen to be an adjoint. Let us elaborate a bit on this. Indeed, it is an empirical 
observation that all perfect matchings should contain the same number of bifundamental fields. This can be traced back to the fact that the rows of the $K$ matrix should add up to the
same number. In this case, the $K$ matrix calculated from $T$ in eq.(\ref{Tmatrix}) has six columns, and a seventh one is added to force this constraint. This can then be seen
to result in a consistent perfect matching matrix, which is given in eq.(\ref{Tmatrix}). \footnote{For theories that do not have a tiling description, added fields in the $K$ matrix
need not be adjoints, as we will see later.}

\begin{figure}[t!]
\begin{minipage}[b]{0.5\linewidth}
\centering
\includegraphics[width=3in,height=2.5in]{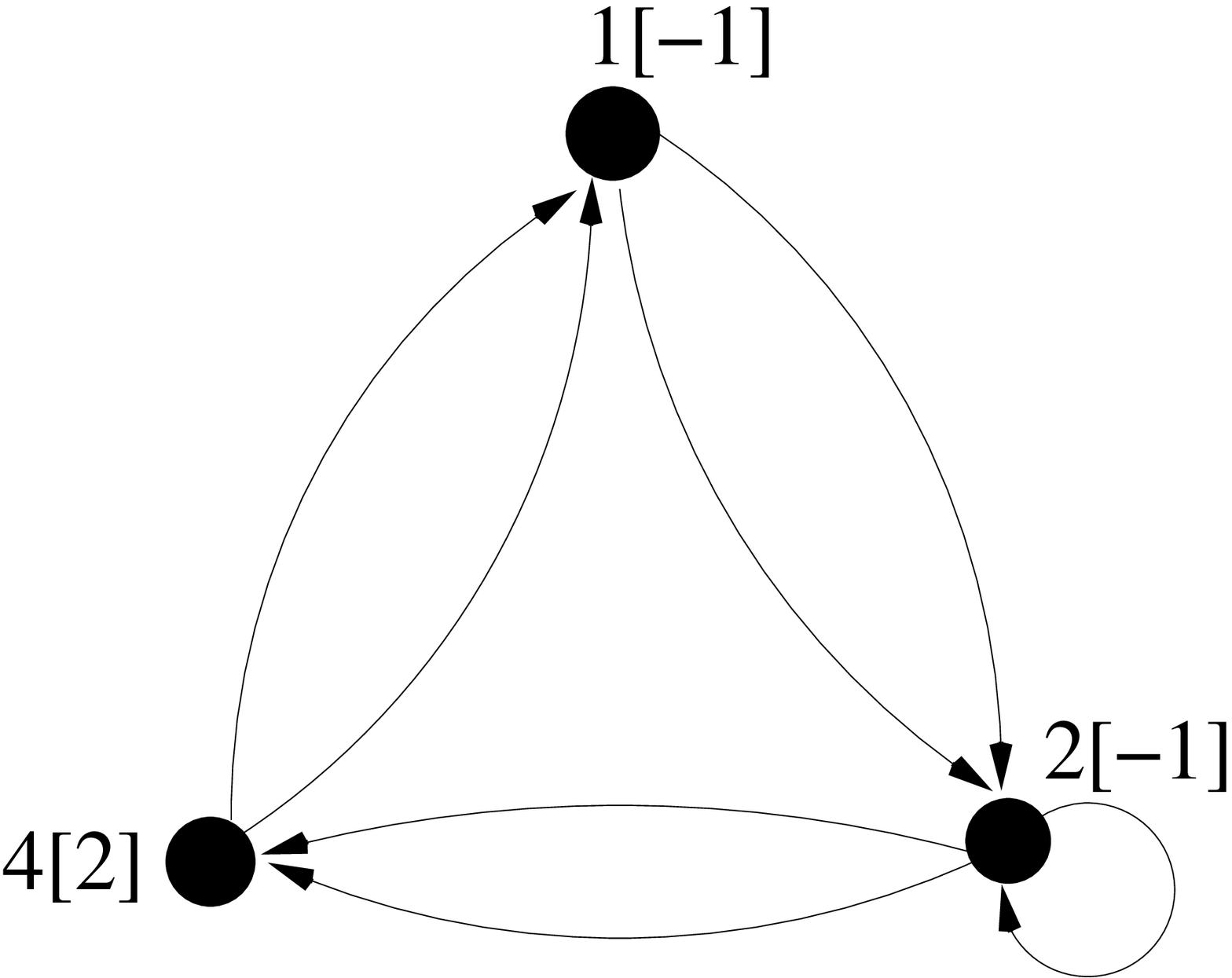}
\caption{Quiver upon Higgsing the ${\mathcal D_1}$ by a single field $X_{23}^2$}
\label{d1higa}
\end{minipage}
\hspace{0.6cm}
\begin{minipage}[b]{0.5\linewidth}
\centering
\includegraphics[width=3in,height=2.5in]{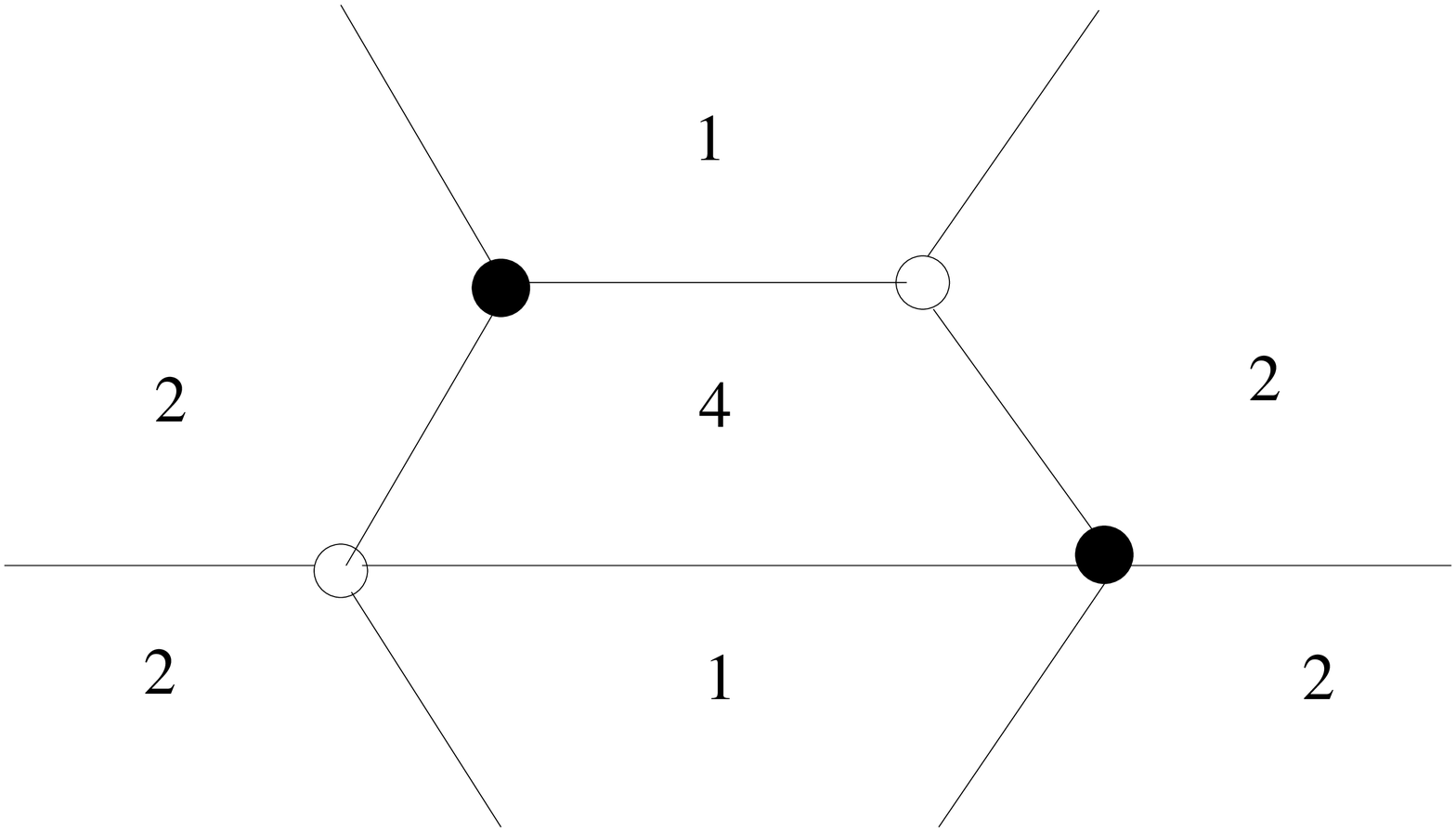}
\caption{Brane tiling for the ${\mathcal D_1}$ theory Higgsed by a single field $X_{23}^2$ }
\label{d1higtileb}
\end{minipage}
\end{figure}

From the tiling picture presented in fig.(\ref{d1higa}), the superpotential of this theory is given by
\begin{equation}
W= \phi\left(X_{24}^2X_{41}^1X_{12}^1 - X_{24}^1X_{41}^1X_{12}^2\right) + X_{12}^2X_{24}^1X_{41}^2 - X_{12}^1X_{24}^2X_{41}^2
\end{equation}
The charge matrix and the consequent toric data are given by
\begin{equation}
Q = 
\begin{pmatrix}
0&1&1&0&-1&-1\\
0&0&0&1&1&-2
\end{pmatrix},~~
{\cal T} = 
\begin{pmatrix}
1&0&0&1&-1\\
0&1&0&-1&0\\
0&0&1&0&0
\end{pmatrix}
\end{equation}
The quiver diagram is the same as that of phase II of $\BC^2/\BZ_2\times \BC^2$ \cite{hanhigg} and is presented in fig.(\ref{d1higa}), and the corresponding brane
tiling is given in fig.(\ref{d1higtileb}). Note that this theory is not that of a cone over a Fano 3-fold.

Now we move on to more complicated cases. We focus on examples where the
Higgsed theory is again a Fano 3-fold. We begin by studying the example of Higgsing the ${\mathcal F_2}$ theory to ${\mathcal E_1}$. 
The ${\mathcal F_2}$ theory has 6 gauge groups and 12 chiral multiplates, with the CS levels being $\vec{k} = \left( 0, -1, 0, -1,1, 1\right)$. The superpotential is
given by
\begin{eqnarray}
W &=& X_{12} X_{23}^{1} X_{31}^{2} + X_{34} X_{42}^{1} X_{23}^{2}  + X_{26}X_{63} X_{31}^{1} X_{15} X_{54} X_{42}^{2}\nonumber\\
&-& X_{12} X_{23}^{2} X_{31}^{1} - X_{34}X_{42}^{2} X_{23}^{1}  - X_{26} X_{63} X_{31}^{2} X_{15} X_{54} X_{42}^{1}
\end{eqnarray}
The perfect matching matrix constructed from the superpotential reads
\begin{equation}
P = 
\begin{pmatrix}
& p_{1} & p_{2} & p_{3} & p_{4} &p_{5} & p_{6}& p_{7} & p_{8} & p_{9}& p_{10}& p_{11} & p_{12}  \\ 
X_{23}^{1} & 1 & 0 & 1 & 0 & 1 & 0 & 1 & 0 & 1 & 0 & 0 & 0 \\ 
X_{23}^{2} & 0 & 1 & 1 & 0 & 1 & 0 & 1 &0 & 1 & 0 & 0 & 0  \\ 
X_{12} & 0 & 0 & 0 & 1 & 0 & 1 & 0 & 1 & 0 & 1 & 1 & 0 \\
X_{34} & 0 & 0 & 0 & 1 & 0 & 1 & 0 & 1 & 0 & 1 & 0 & 1\\
X_{42}^{1}& 1 & 0 & 0 & 0 & 0 & 0 & 0 & 0 & 0 & 0 & 1 & 0\\ 
X_{42}^{2} & 0 & 1 & 0 & 0 & 0 & 0 & 0  & 0 & 0 & 0 & 1 & 0 \\ 
X_{31}^{1} &  1 & 0 & 0 & 0 & 0 & 0 & 0 & 0 & 0 & 0 & 0 & 1\\ 
X_{31}^{2}& 0 & 1 & 0 & 0 & 0 & 0 & 0 & 0 & 0 & 0 & 0 & 1\\ 
X_{15} & 0 & 0 & 1 & 0 & 0 & 0 & 0 & 0 & 0 & 1 & 0 & 0 \\ 
X_{63} & 0 & 0 & 0 & 1 & 0 & 0 & 0 & 0 & 1 & 0 & 0 & 0 \\
X_{54} & 0 & 0 & 0 & 0 & 1 & 0 & 0 & 1 & 0 & 0 & 0 & 0 \\ 
X_{26} & 0 & 0 & 0 & 0 & 0 & 1 & 1 & 0 & 0 & 0 & 0 & 0
\end{pmatrix}
\end{equation}
\begin{figure}[t!]
\begin{minipage}[b]{0.5\linewidth}
\centering
\includegraphics[width=3in,height=2.5in]{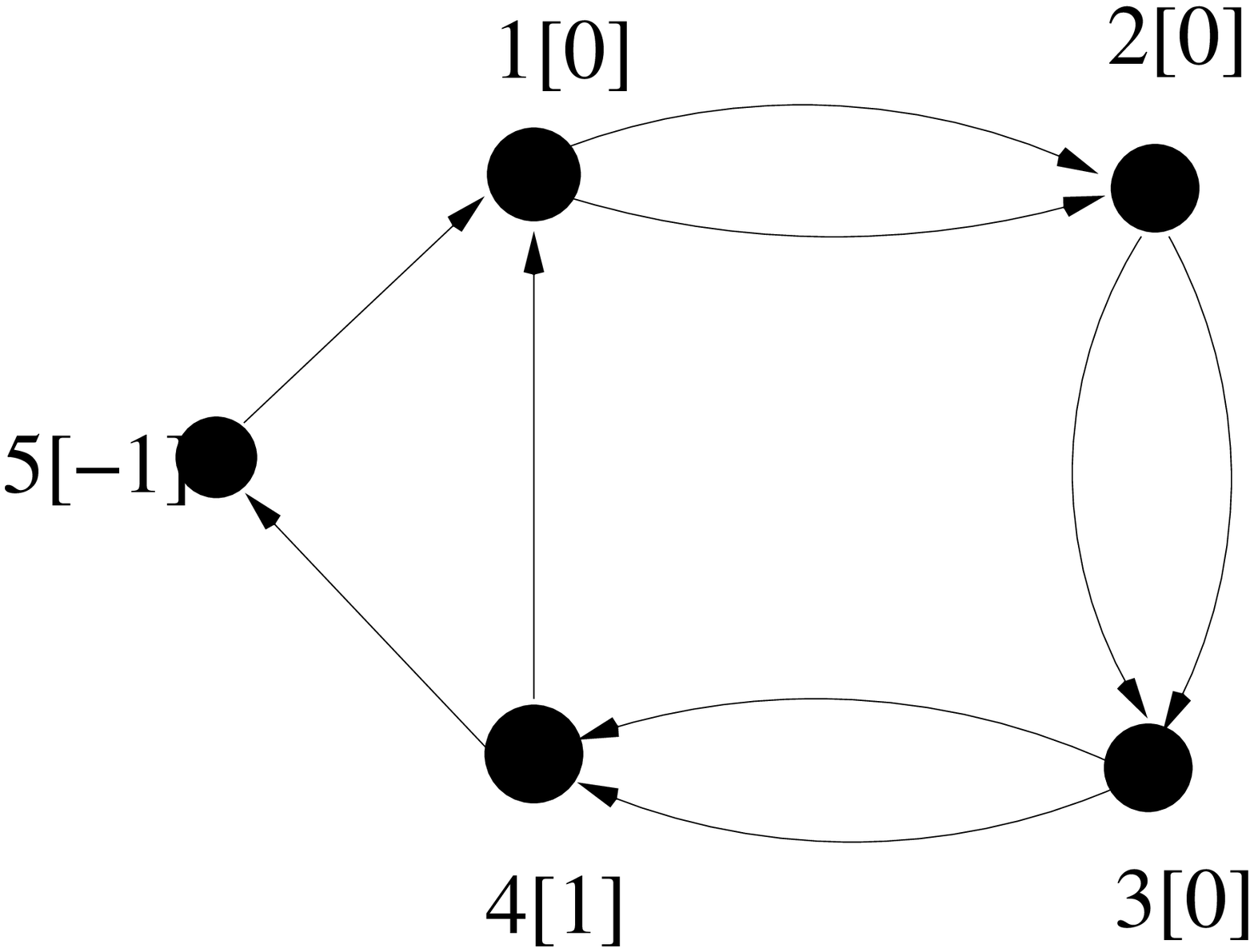}
\caption{Higgsing the ${\mathcal F_1}$ by a single field $X_{46}$}
\label{higf11}
\end{minipage}
\hspace{0.6cm}
\begin{minipage}[b]{0.5\linewidth}
\centering
\includegraphics[width=3in,height=2.5in]{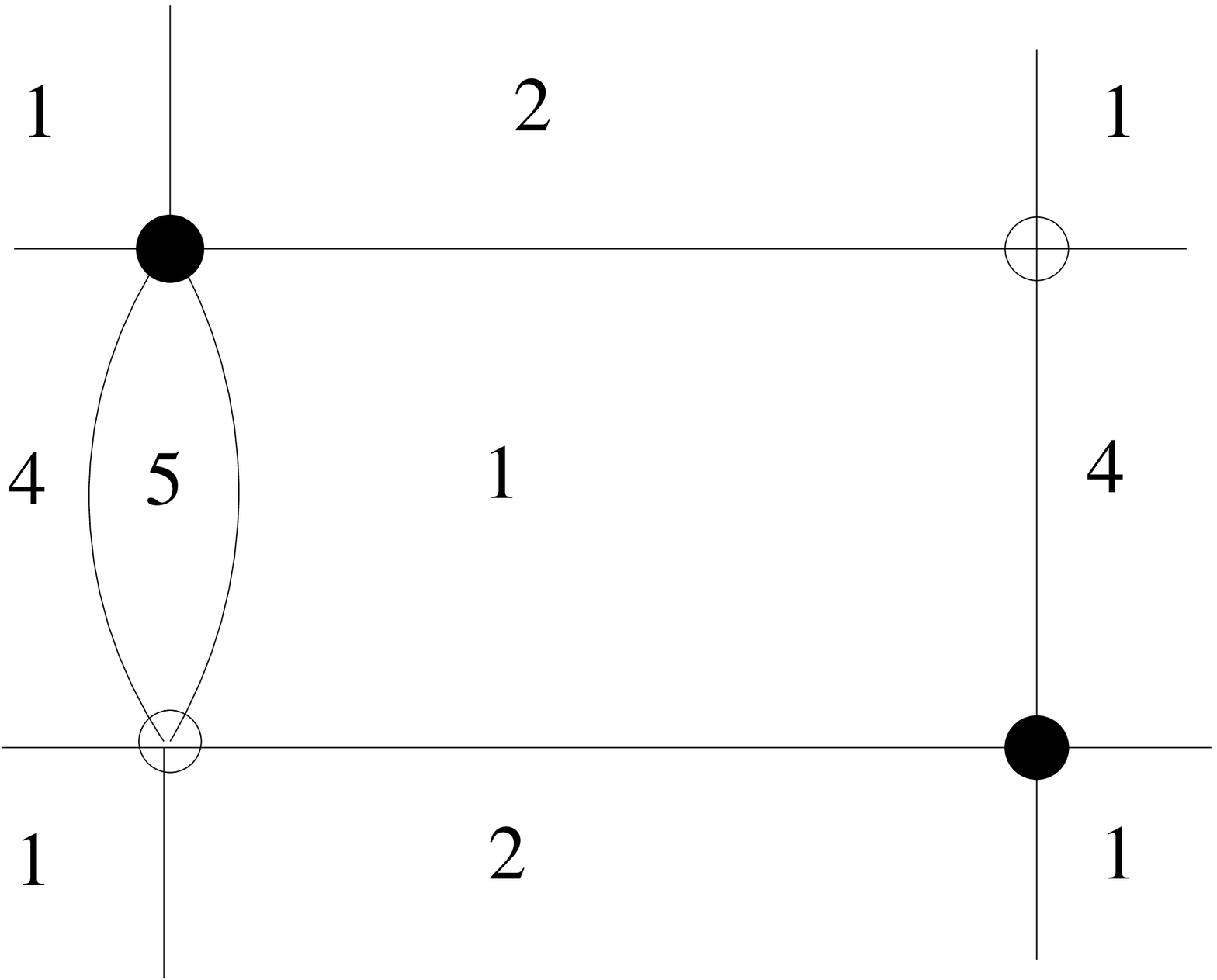}
\caption{Brane tiling for the ${\mathcal F_1}$ theory Higgsed by a single field $X_{46}$ }
\label{higtilingf11}
\end{minipage}
\end{figure}
We choose to remove the points $p_4$ and $p_9$ from the parent theory. To this effect, we Higgs the theory by giving a vev to the field $X_{63}$, i.e remove the 10th row, 4th 
and the 9th columns from the matching matrix P. The resulting $T$ and the dual $K$ matrices are 
\begin{equation}
T = 
{\small
\begin{pmatrix}
1 & 0 & 0  & 0 & 0 & 0 & 0  & 0 & 0 & 1 \\  
1 & 0 & 0  & 0 & 0 & 0 & 0  & 0 & 1 & 0  \\  
0 & 0 & 1  & 0 & 0 & 0 & 0  & 1 & 0 & 0\\  
0 & 0 & 0  & 1 & 0 & 0 & 1  & 0 & 0 & 0   \\  
1 & 0 & 1  & 1 & 0 & 1 & 0  & 0 & 0 & 0\\  
-1 & 0 & -1  & -1 & 1 & 0 & 0  & 0 & 0 & 0 \\  
-1 & 1 & 0  & 0 & 0 & 0 & 0  & 0 & 0 & 0
\end{pmatrix}},~
K = 
{\small
\begin{pmatrix}
0 & 0 & 0  & 0 & 0 & 0 & 0 & 0 & 1 & 1 & 1 \\ 
 0 & 0 & 0  & 0 & 0 & 1 & 1  & 1 & 0 & 0 & 0 \\  
 0 & 0 & 0 & 0 & 1 & 0 & 0  & 1 & 0 & 0 & 1\\  
 0 & 0 & 0  & 1 & 0 & 0 & 0  & 1 & 0 & 0 & 1 \\  
1 & 1 & 1  & 0 & 0 & 0 & 0  & 0 & 0 & 0 & 0\\  
0 & 0 & 1  & 0 & 0 & 0 & 0  & 1 & 0 & 0 & 1  \\  
0 & 1 & 0  & 0 & 0 & 0 & 1  & 0 & 0 & 1 & 0
\end{pmatrix}}
\end{equation}
The reduced matching matrix can be calculated, using the above.
The superpotential of the new theory is obtained from the $K$ matrix as
\begin{equation}
W = X_{3} X_{4} X_{5} X_{6} X_{10} + X_{1} X_{7} X_{11} + X_{2} X_{8} X_{9} -
X_{3} X_{4} X_{5} X_{7} X_{9}-X_{1} X_{8} X_{10}-X_{2} X_{6} X_{11}
\end{equation}
Using the perfect matching matrix and the superpotential, we can construct the brane tiling. The daughter theory has 5 gauge groups and its CS levels are,
$\vec{k} = \left( 0, -1, 1, -1,1\right)$. From the tiling picture, the quiver charge matrix is obtained to be 
\begin{equation}
d = 
\begin{pmatrix}
&X_{1} & X_{2} & X_{3} & X_{4} & X_{5} & X_{6} & X_{7} & X_{8} & X_{9} & X_{10} & X_{11} \\ 
G_{1} &  0 & 0 & 0  & 1 & 0 & 0 & 0  & 1 & -1 & -1 & 0   \\ 
G_{2} &  1 & 1 & 0 & 0 & 1 & -1 & -1  & -1 & 0 & 0 & 0\\ 
G_{3} & -1 & -1 & 0  & 0 & -1 & 0 & 0  & 0 & 1 & 1 & 1  \\ 
G_{4} &  0 & 0 & -1  & 0 & 0 & 1 & 1  & 0 & 0 & 0 & -1\\ 
G_{5} & 0 & 0 & 1  & -1 & 0 & 0 & 0  & 0 & 0 & 0 & 0
\end{pmatrix}
\end{equation}
where $G_i, i=1\cdots 5$ denote the five gauge groups. The face symmetries of the daughter theory are found to be
\begin{eqnarray}
F_{1} &=& X_{12}+ X_{15}-X_{31}^{1}-X_{31}^{2} = p_{7}-p_{10}\nonumber\\
F_{2} &=& X_{23}^{1}+ X_{23}^{2}+ X_{23}^{3}-X_{12}-X_{42}^{1}-X_{42}^{2} =p_{3}-p_{9} \nonumber\\
F_{3} &=& X_{31}^{1}+ X_{31}^{2}+ X_{34}-X_{23}^{1}-X_{23}^{2}- X_{23}^{3} = p_{10}-p_{3}\nonumber\\
F_{4} &=& X_{42}^{1}+ X_{42}^{2}-X_{34}-X_{54} = p_{4}+ p_{9}-p_{6}-p_{7} \nonumber\\
F_{5}&=& X_{54}-X_{15} = p_{6}-p_{4}
\end{eqnarray}
Now we can calculate the baryonic charge matrix 
\begin{equation}
Q_D=
\begin{pmatrix}
0 & 0  & 0 & 0 & 0 & 0  & 1 & 0 & 0 & -1 \\   
0 & 0 & 0 & 0 & 0 & 0  & 0 & 0 & -1 & 1\\  
0 & 0  & 1 & -1 & 0 & 1  & 0 & 0 & -1 & 0
\end{pmatrix}
\end{equation}
The toric data can now be determined from the kernel of the matrix obtained by concatenating the baryonic charge matrix with the
kernel of the reduced matching matrix, and is given by
\begin{equation}
{\cal T} = 
\begin{pmatrix}
 -1 & 1  & 0 & 0 & 0 & 0 & 0 & 0 & 0 & 0 \\  
 1 & 0  & 1 & 1 & -1 & 0  & 0 & 0 & 0 & 0 \\   
 0 & 0 & 1 & 0 & -1 & -1  & 0 & 1 & 0 & 0 
\end{pmatrix}
\end{equation}
This is the theory ${\mathcal E_1}$ of \cite{fanohan}. 

The next example that we study is the Higgsing of the theory ${\mathcal D_2}$, which is a theory of four gauge groups and ten chiral multiplets. 
The superpotential of the theory is given by 
\begin{eqnarray}
W &=& X_{31}^{1} X_{12}^{2} X_{23}^{3} + X_{12}^{1} X_{23}^{2} X_{31}^{3}  + X_{23}^{1} X_{31}^{2} X_{14} X_{42}\nonumber\\
&-&X_{31}^{2} X_{12}^{1} X_{23}^{3} - X_{12}^{2}X_{23}^{1} X_{31}^{3}  - X_{23}^{2} X_{31}^{1} X_{14} X_{42}
\end{eqnarray}
It can be seen that giving a vev to the field $X_{14}$ gives rise to the theory ${\mathcal B_4}$ with toric data
\begin{equation}
{\cal T} = 
\begin{pmatrix}
1 & -1  & 0 & 0 & 0 & 0 \\ 
-1 & 0 & 1 & 0  & 0 & 0  \\ 
0 & 0  & 0 & -1 & 0 & 1
\end{pmatrix}
\end{equation}
An entirely similar analysis can be carried out to Higgs the Fano 3-fold theory ${\mathcal E_3}$ to the theory ${\mathcal C_3}$ and 
the theory ${\mathcal C_4}$ to ${\mathcal B_4}$. 

\begin{figure}[t!]
\begin{minipage}[b]{0.5\linewidth}
\centering
\includegraphics[width=3in,height=2.5in]{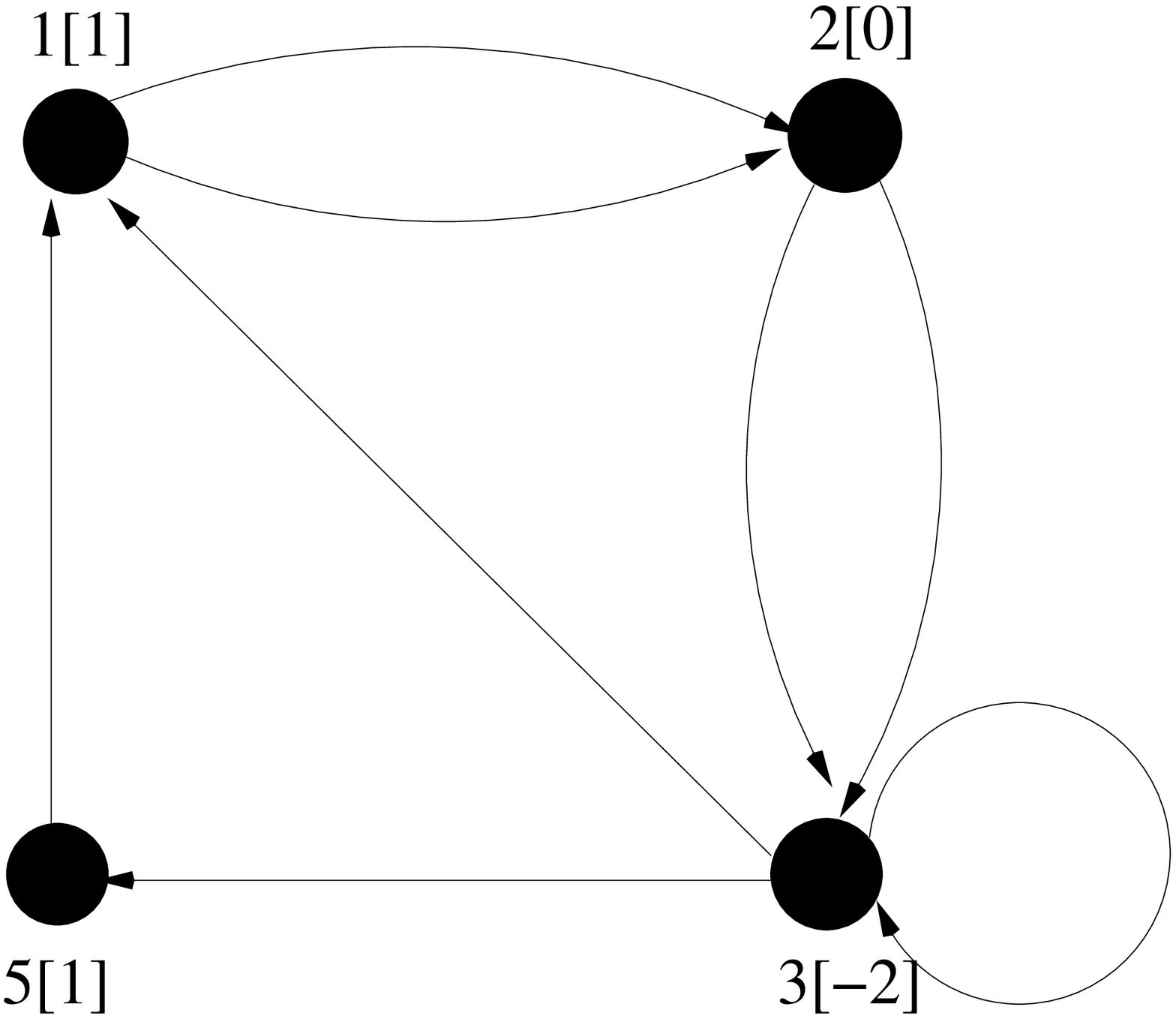}
\caption{Higgsing the ${\mathcal E_2}$ by the field $X_{34}^2$}
\label{hige21}
\end{minipage}
\hspace{0.6cm}
\begin{minipage}[b]{0.5\linewidth}
\centering
\includegraphics[width=3in,height=2.5in]{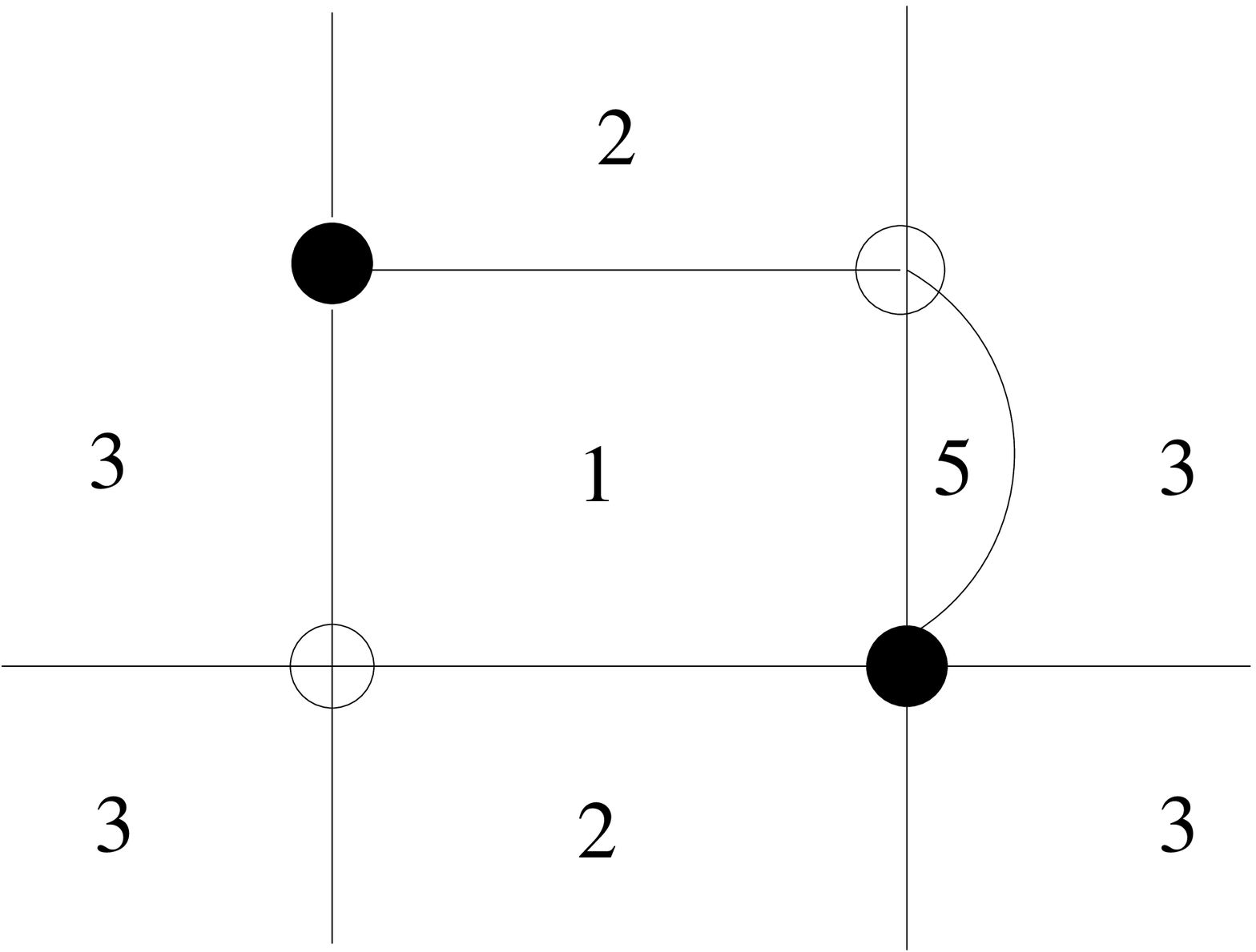}
\caption{Brane tiling for the ${\mathcal E_2}$ theory Higgsed by the field  $X_{34}^2$}
\label{higtilinge21}
\end{minipage}
\end{figure}

Let us now point out a couple of non-Fano theories which appear on Higgsing
Fano 3-folds. As a first example, we consider the Higgsing of the theory ${\mathcal F_1}$ by a single field, say $X_{46}$, in the notation of \cite{fanohan}.
The resulting theory has a superpotential 
\begin{equation}
W = X_{12}^{1} X_{23}^{1} X_{34}^{2} X_{45} X_{51} -  X_{12}^{2} X_{23}^{1}
X_{34}^{1} X_{45} X_{51} + X_{12}^{2} X_{23}^{2} X_{34}^{1} X_{41}-X_{12}^{1}
X_{23}^{2} X_{34}^{1} X_{41}
\end{equation}
The quiver diagram and the brane tiling picture is presented in fig.(\ref{higf11}) and fig.(\ref{higtilingf11}) where we have also indicated the CS levels of the Higgsed theory.
The toric data can be shown to be 
\begin{equation}
{\cal T}=
\begin{pmatrix} 
-1 & 1 & 0 & 0 & 0 & 0 & 0 & 0 & 0 & 0 \\ 
0 & 0 & 1 & 0 & 0 & 0 & 0 & 0 & 1 & 0 \\  
0 & 0 & -1 & 1 & -1 & 0 & 0 & 0 & 0 & 0  
\end{pmatrix}
\end{equation}
This is an interesting theory with the toric data corresponding to $\BP^1 \times {\rm SPP}$ where the well known SPP (suspended pinch point) singularity arises as 
a partial resolution of the orbifold $\BC^3/\BZ_2\times\BZ_2$.
Finally, let us consider Higgsing the theory ${\mathcal E_2}$ by a single field which we choose to be $X_{34}^{2}$ in the notation of \cite{fanohan}.
The resulting superpotential is
\begin{equation}
W = X_{33} X_{35} X_{51} X_{12}^{1} X_{23}^{1} -  X_{33} X_{12}^{1} X_{23}^{2}
X_{31} + X_{12}^{2} X_{23}^{2} X_{31}-X_{12}^{2} X_{23}^{1} X_{35} X_{51}
\end{equation}
The quiver diagram and the brane tiling for the resulting theory is given in fig.(\ref{hige21}) and fig.(\ref{higtilinge21}). This corresponds to the data
\begin{equation}
{\cal T}=
\begin{pmatrix} 
1&0&0&0&0&0&0&0\\
0&1&0&0&-1&-1&0&0\\
0&0&1&1&0&1&-1&0
\end{pmatrix}
\end{equation}
We end by summarising the results presented in this section. Here, we have explicitly studied the Higgsing of complex cones over smooth Fano 3-folds using the
method of \cite{tapo1}. We have established the Higgsing of ${\mathcal F}_2$ to ${\mathcal E}_1$, ${\mathcal D}_2$ to ${\mathcal B}_4$, 
${\mathcal E}_3$ to ${\mathcal C}_3$ and ${\mathcal C}_4$ to ${\mathcal B}_4$. These are not surprising, given the matter content, superpotential and
the CS levels of the theories as given in \cite{hanhigg}. We have also studied the Higgsing of the theories ${\mathcal D}_1$, ${\mathcal F}_1$ (which gave the 
interesting theory $\BP^1\times {\rm SPP}$) and ${\mathcal E}_2$ and established that
at scales much below the Higgs scale, these should flow to CY 4-folds which are not cones over smooth Fano varieties. \footnote{These theories in many cases seem to
correspond to $\BP^1$ bundles over surfaces which are not del Pezzo.} These are the main results of this
section. A couple of words are however in order. 

As mentioned previously, in three complex dimensions, cones over Fano 2-folds are related to each other by Higgsing and unHiggsing, 
which correspond to the blowup or blowdown of $\BP^1$s at generic or non-generic points. Even for cones over Fano 3-folds, similar results exist, and it should be 
possible to find these in the gauge theory picture. It is, for example, known that starting from the variety ${\mathcal F}_1$, i.e $\BP^1 \times dP_3$, we can, by blowing
down a divisor (there are six equivalent choices), reach the variety ${\mathcal E}_3$, and that a further blowdown should result in ${\mathcal C}_3$ or  ${\mathcal C}_4$.
By a Higgsing procedure, we have established the last one. There should thus be different phases of ${\mathcal F}_1$ or ${\mathcal E}_3$ which should correspond
to the other two. Mapping out the entire structure of Higgsing Fano 3-folds which will reproduce all the known relations between these is a daunting task, and it should
be studied further. 

\section{UnHiggsing Fano 3-Folds}

In this section, we study the unHiggsing procedure for toric Fano 3-folds, where we add fields to a given theory to obtain a different gauge theory. If the resulting variety is
a toric CY, it should satisfy appropriate convexity conditions. These are expected to give rise to new theories, just as unHiggsing of Fano 2-folds 
give rise to pseudo del Pezzo varieties. UnHiggsing M2-brane theories was systematically explored in \cite{bhs} to which we refer the reader for more details.
We remind the reader that unHiggsing the theory introduces new gauge groups, and if we introduce a bifundamental $X_{ij}$, say, where $j$ is the new gauge
group introduced, then the CS levels of $i$ and $j$ are chosen so that they add up to the original CS assignment for the group $i$. One can also add three or
five fields in order to add a single gauge group to the original theory. The superpotentials are appropriately modified in these cases, as discussed in \cite{bhs}.
We will mostly consider cases where we add one or three fields to the original quiver. We will also discuss a generalisation of the method of \cite{bhs} where we 
perform an unHiggsing procedure where an adjoint field is present in the original theory. 

We begin with the study of the ${\mathcal F_1}$ theory. This model has 6 gauge groups and 10 chiral multiplates. 
The CS levels are, $\vec{k} = \left( 0, 0, 0, 0,-1, 1\right)  $ The superpotential for this model is given by,
\begin{equation}
W= X_{12}^{1} X_{23}^{1} X_{34}^{2} X_{45} X_{51} +X_{12}^{2} X_{23}^{2}
X_{34}^{1} X_{46} X_{61} -X_{12}^{2} X_{23}^{1} X_{34}^{1} X_{45} X_{51}-X_{12}^{1}
X_{23}^{2} X_{34}^{2} X_{46} X_{61}
\end{equation}
The toric data for this Fano variety is given by 
\begin{equation}
{\cal T} = 
\begin{pmatrix}
1 & -1 & 0 & 0 & 0 & 0 & 0 & 0 & 0 & 0 & 0 & 0 & 0 \\
0 & 0 & 1 & -1 & 1 & -1 & 0 & 0 & 0 & 0 & 0 & 0 & 0 \\ 
0 & 0 & 0 & 0 & 1 & -1 & -1 & 1 & 0 & 0 & 0 & 0 & 0
\end{pmatrix}
\end{equation}

We first proceed by unHiggsing with a single field, say $X_{67}$. The theory obtained will have 7 gauge groups and 11 fields.
We choose the new CS levels to be, $\vec{k} = \left( 0, 0, 0, 0,-1, 1,0 \right)$. The superpotential can be shown to be modified as
\begin{eqnarray}
W &=& X_{12}^{1} X_{23}^{1} X_{34}^{2} X_{45} X_{51} + X_{12}^{2} X_{23}^{2}X_{34}^{1} X_{46} X_{67} X_{71} \nonumber\\
&-& X_{12}^{2} X_{23}^{1} X_{34}^{1} X_{45}X_{51} - X_{12}^{1} X_{23}^{2} X_{34}^{2} X_{46} X_{67} X_{71}
\end{eqnarray}
From the superpotential we construct the matching matrix using standard methods.It contains 16 perfect matchings with each involving 2 bifundamental fields.
The explicit form of the perfect matching matrix is 
\begin{equation}
P = 
\begin{pmatrix}
& p_{1} & p_{2} & p_{3} & p_{4} & p_{5} & p_{6} & p_{7} & p_{8} & p_{9} & p_{10} & p_{11} & p_{12} &
p_{13} & p_{14} & p_{15} & p_{16} \\ 
X_{12}^{1} & 1 & 1 & 0 & 0 & 0 & 0 & 0 & 0 & 0 & 0 & 0 & 0 & 0 & 0 & 0 & 0\\ 
X_{12}^{2} & 1 & 0 & 1 & 0 & 0 & 0 & 0 & 0 & 0 & 0 & 0 & 0 & 0 & 0 & 0 & 0\\
X_{23}^{1} & 0 & 0 & 0 & 0 & 1 & 1 & 1 & 1 & 0 & 0 & 0 & 0 & 0 & 0 & 0 & 0\\
X_{23}^{2} & 0 & 0 & 0 & 0 & 1 & 0 & 0 & 0 & 1 & 0 & 0 & 0 & 1 & 0 & 0 & 0\\ 
X_{34}^{1} & 0 & 1 & 0 & 1 & 0 & 0 & 0 & 0 & 0 & 0 & 0 & 0 & 0 & 0 & 0 & 0\\ 
X_{34}^{2} & 0 & 0 & 1 & 1 & 0 & 0 & 0 & 0 & 0 & 0 & 0 & 0 & 0 & 0 & 0 & 0\\ 
X_{45} & 0 & 0 & 0 & 0 & 0 & 0 & 0 & 0 & 1 & 1 & 1 & 1 & 0 & 0 & 0 & 0\\ 
X_{46} & 0 & 0 & 0 & 0 & 0 & 1 & 0 & 0 & 0 & 1 & 0 & 0 & 0 & 1 & 0 & 0\\
X_{51} & 0 & 0 & 0 & 0 & 0 & 0 & 0 & 0 & 0 & 0 & 0 & 0 & 1 & 1 & 1 & 1\\ 
X_{67} & 0 & 0 & 0 & 0 & 0 & 0 & 1 & 0 & 0 & 0 & 1 & 0 & 0 & 0 & 1 & 0\\ 
X_{71} & 0 & 0 & 0 & 0 & 0 & 0 & 0 & 1 & 0 & 0 & 0 & 1 & 0 & 0 & 0 & 1
\end{pmatrix}
\end{equation}
It can also be checked that there are seven face symmetries of this model, given by the equations 
\begin{eqnarray}
F_{1} &=& X_{12}^{1}+ X_{12}^{2}-X_{51}-X_{71} = p_{1}-p_{16},~~F_{2} = X_{23}^{1}+ X_{23}^{2}-X_{12}^{1}-X_{12}^{2} = p_{5}-p_{1} \nonumber\\
F_{3} &=& X_{34}^{1}+ X_{34}^{2}-X_{23}^{1}-X_{23}^{2} = p_{4}-p_{5},~~F_{4} = X_{45}+ X_{46}-X_{34}^{1}-X_{34}^{2} = p_{10}-p_{4} \nonumber\\
F_{5} &=& X_{51}-X_{45} = p_{16}-p_{12},~~F_{6} = X_{67}-X_{46} = p_{11}-p_{10} \nonumber\\
F_{7} &=& X_{71}-X_{67} = p_{12}-p_{11} 
\end{eqnarray}
where the second equality expresses the face symmetries in terms of the perfect matchings. 
The charge matrix $Q_D$ can now be obtained from these face symmetries as
\begin{equation}
Q_D=
\begin{pmatrix}
p_{1} & p_{2} & p_{3} & p_{4} & p_{5} & p_{6} & p_{7} & p_{8} & p_{9} & p_{10} & p_{11} & p_{12} & p_{13} & p_{14} & p_{15} & p_{16} \\  
1 & 0 & 0 & 0 & 0 & 0 & 0 & 0 & 0 & 0 & 0 & 0 & 0 & 0 & 0 & -1\\
-1 & 0 & 0 & 0 & 1 & 0 & 0 & 0 & 0 & 0 & 0 & 0 & 0 & 0 & 0 & 0\\ 
0 & 0 & 0 & 1 & -1 & 0 & 0 & 0 & 0 & 0 & 0 & 0 & 0 & 0 & 0 & 0\\ 
0 & 0 & 0 & -1 & 0 & 0 & 0 & 0 & 0 & 1 & 0 & 0 & 0 & 0 & 0 & 0\\ 
0 & 0 & 0 & 0 & 0 & 0 & 0 & 0 & 0 & 0 & -1 & 1 & 0 & 0 & 0 & 0
\end{pmatrix}
\end{equation}
The matrix $Q_F$ is obtained from the kernel of the perfect matching matrix. The toric data (after a suitable $GL\left(4,Z\right)$ transformation can be cast 
in the form 
\begin{equation}
{\cal T} = 
\begin{pmatrix}
0 & -1 & 1 & 0 & 0 & 0 & 0 & 0 & 0 & 0 & 0 & 0 & 0 & 0 & 0 & 0\\
0 & 0 & 0 & 0 & 0 & 1 & 0 & 0 & -1 & 0 & -1 & -1 & 0 & 1 & 0 & 0\\ 
0 & 0 & 0 & 0 & 0 & 1 & 1 & 1 & -1 & 0 & 0 & 0 & -1 & 0 & 0 & 0
\end{pmatrix}
\end{equation}
\begin{figure}[t!]
\begin{minipage}[b]{0.5\linewidth}
\centering
\includegraphics[width=3in,height=2.5in]{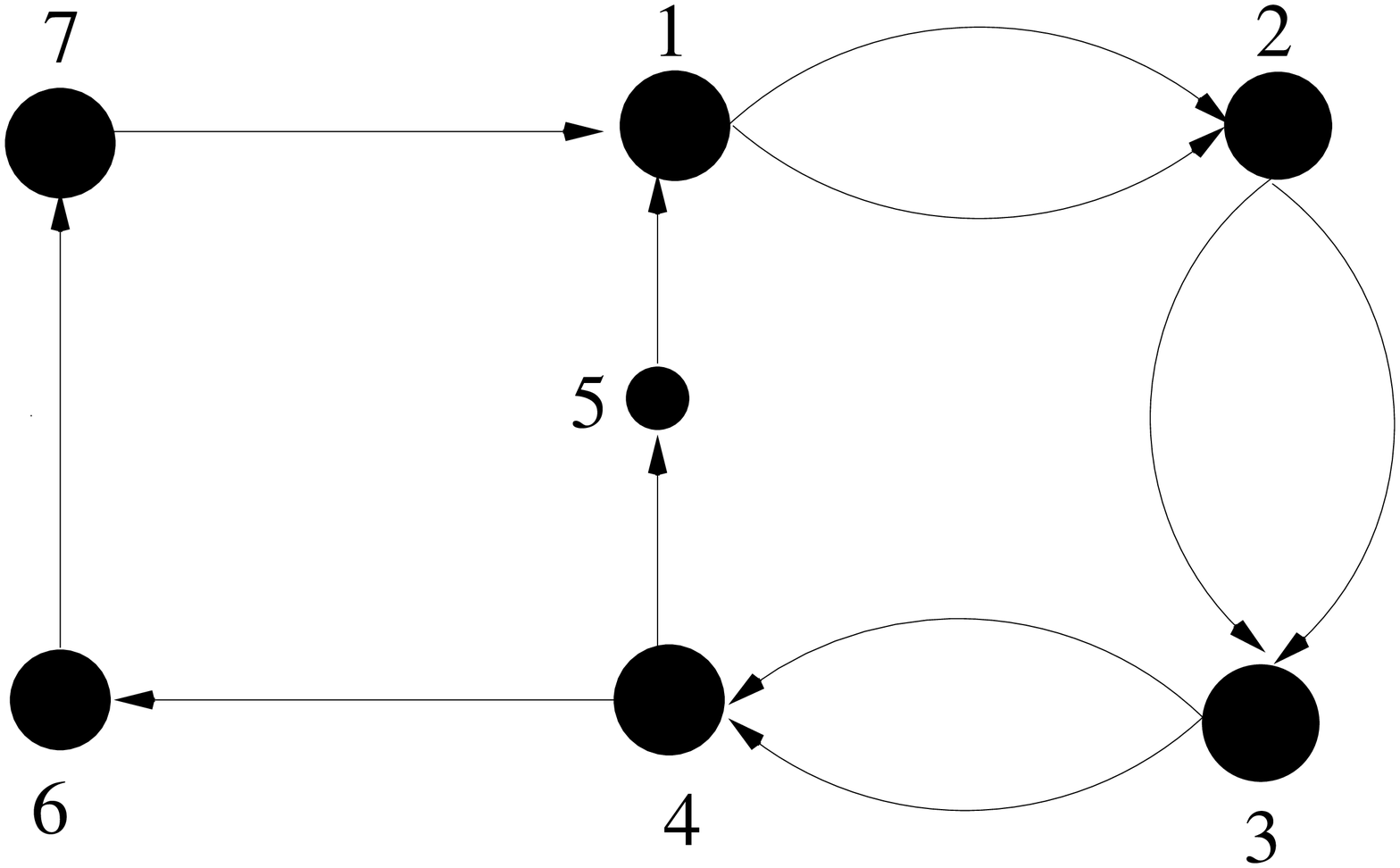}
\caption{UnHiggsing the ${\mathcal F_1}$ by a single field $X_{67}$}
\label{unhigf11}
\end{minipage}
\hspace{0.6cm}
\begin{minipage}[b]{0.5\linewidth}
\centering
\includegraphics[width=3in,height=2.5in]{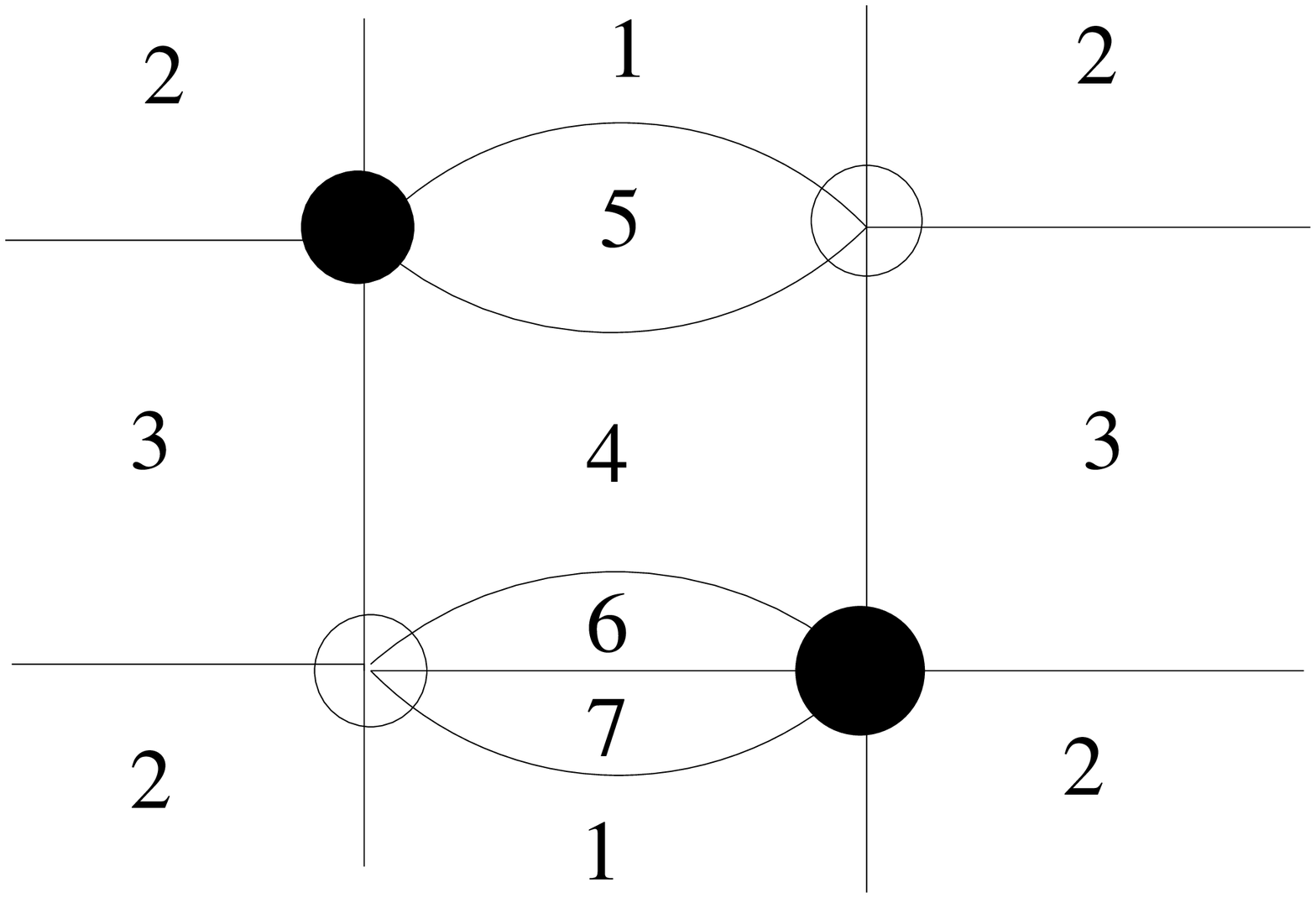}
\caption{Brane tiling for the ${\mathcal F_1}$ theory unHiggsed by a single field $X_{67}$ }
\label{tilingf11}
\end{minipage}
\end{figure}

Note that the toric data is the same as that of the theory ${\mathcal F_1}$, with different multiplicities. In particular, just as in the Fano 2-fold examples,
the external points have acquired multiplicities. UnHiggsing, therefore, has given back the same theory, although with a different matter content and
superpotential. The quiver diagram and the brane tiling tiling picture for the unHiggsed theory are presented in fig.(\ref{unhigf11}) and fig.(\ref{tilingf11}). 
We could, of course, choose a different set of CS levels in the unHiggsed theory. The superpotential, perfect matching matrix and hence 
the $Q_{F}$ matrices will remain the same, but $ Q_{D} $ and hence hence the toric data will be modified. Suppose we choose the CS levels in the
unHiggsed theory as $\vec{k} = \left( 0, 0, 0, 0,-1, 2,-1 \right)$. This is also a valid choice of CS levels, given the discussion in the beginning of this section. 
We find that in this case, the matrix of baryonic symmetries is given by
\begin{equation}
Q_D=
\begin{pmatrix}
p_{1} & p_{2} & p_{3}& p_{4} & p_{5} & p_{6} & p_{7} & p_{8} & p_{9} & p_{10} & p_{11} & p_{12} & p_{13} & p_{14} & p_{15} & p_{16} \\ 
1 & 0 & 0 & 0 & 0 & 0 & 0 & 0 & 0 & 0 & 0 & 0 & 0 & 0 & 0 & -1\\
-1 & 0 & 0 & 0 & 1 & 0 & 0 & 0 & 0 & 0 & 0 & 0 & 0 & 0 & 0 & 0\\ 
0 & 0 & 0 & 1 & -1 & 0 & 0 & 0 & 0 & 0 & 0 & 0 & 0 & 0 & 0 & 0\\ 
0 & 0 & 0 & -1 & 0 & 0 & 0 & 0 & 0 & 1 & 0 & 0 & 0 & 0 & 0 & 0\\ 
0 & 0 & 0 & 0 & 0 & 0 & 0 & 0 & 0 & -1 & -1 & 2 & 0 & 0 & 0 & 0
\end{pmatrix}
\end{equation}
Now we obtain the toric data (after a $GL\left(4,Z\right)$ transformation) as
\begin{equation}
{\cal T}=
\begin{pmatrix}
0 & -1 & 1 & 0 & 0 & 0 & 0 & 0 & 0 & 0 & 0 & 0 & 0 & 0 & 0 & 0 \\
0 & 0 & 0 & 0 & 0 & -1 & 1 & 0 & 1 & 0 & 2 & 1 & 0 & -1 & 1 & 0\\ 
0 & 0 & 0 & 0 & 0 & -1 & -1 & -1 & 1 & 0 & 0 & 0 & 1 & 0 & 0 & 0
\end{pmatrix}
\end{equation}
This is a toric four-fold that is not a complex cone over a toric Fano 3-fold. As we can see, apart from increasing the multiplicities to certain points,
unhiggsing has also resulted in adding new points to the original toric diagram. This singularity, however, appears to be non-isolated, as there are lattice points
that are internal to one of the edges of the toric diagram. 
\begin{figure}[t!]
\begin{minipage}[b]{0.5\linewidth}
\centering
\includegraphics[width=3in,height=2.5in]{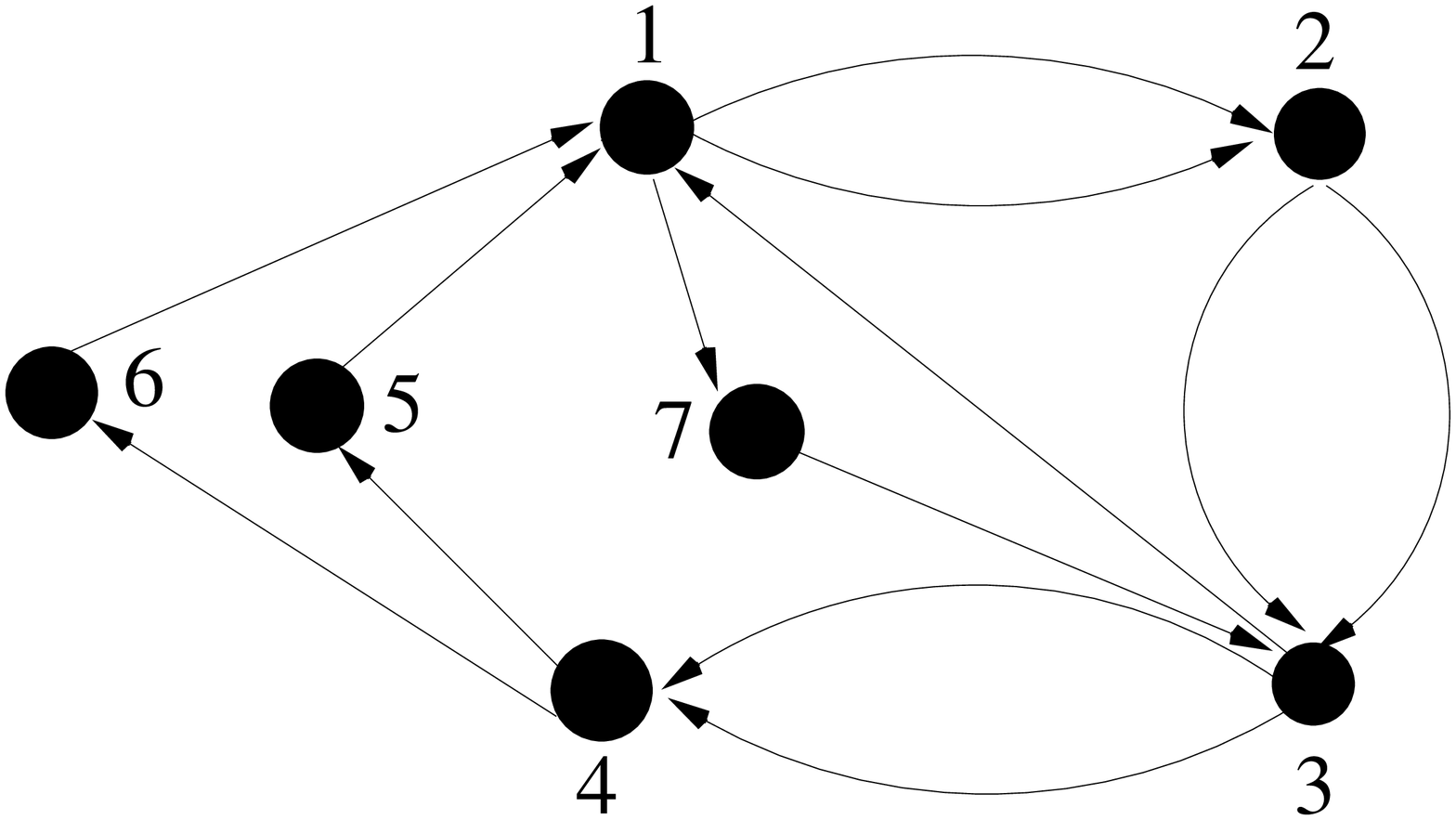}
\caption{Quiver diagram for the ${\mathcal F_1}$ theory unHiggsed by three fields }
\label{unhigf13}
\end{minipage}
\hspace{0.6cm}
\begin{minipage}[b]{0.5\linewidth}
\centering
\includegraphics[width=3in,height=2.5in]{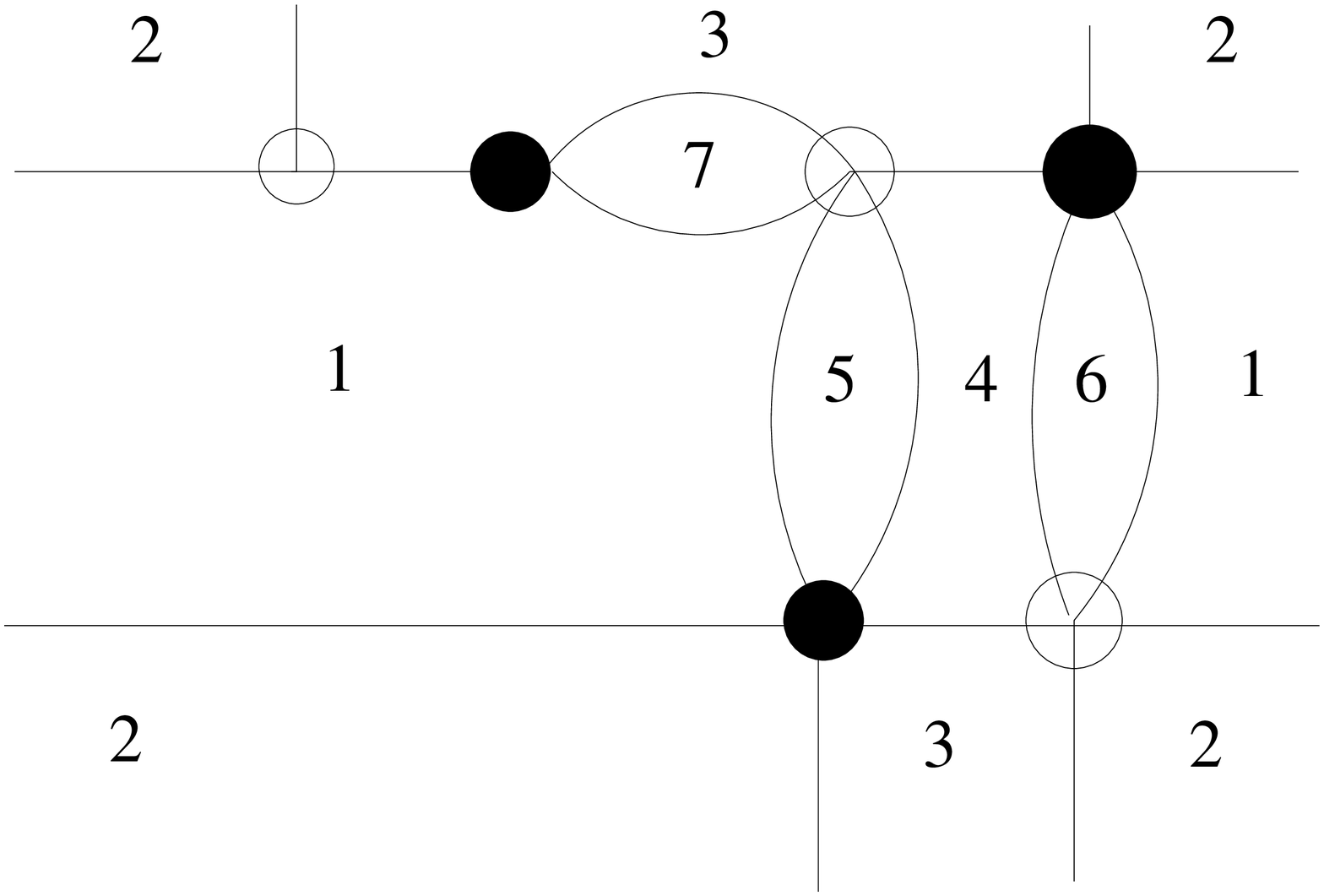}
\caption{Brane tiling for the ${\mathcal F_1}$ theory unHiggsed by three fields. }
\label{tilingf13}
\end{minipage}
\end{figure}

We now consider unHiggsing of the ${\mathcal F_1}$ theory with three fields, such that the final theory  has  7 gauge groups and 13 bifundamental fields.
The CS levels are chosen to be $\vec{k} = \left( -1, 0, 0, 0,-1, 1,1 \right)$, and the superpotential is given by 
\begin{eqnarray}
W &=& X_{12}^{1} X_{23}^{1} X_{34}^{2} X_{45} X_{51}+X_{12}^{2} X_{23}^{2}X_{34}^{1} X_{46} X_{61}+ X_{17} X_{73} X_{31}\nonumber\\
&-& X_{12}^{2} X_{23}^{1} X_{31}-X_{34}^{1} X_{45} X_{51} X_{17} X_{73}-X_{12}^{1} X_{23}^{2} X_{34}^{2} X_{46}X_{61}
\end{eqnarray}
The perfect matching matrix can be readily constructed, given the superpotential. We further note that in this case the face symmetries are obtained as
\begin{eqnarray}
F_{1}&=&  p_{1}-p_{13},~~F_{2}= p_{2}-p_{1},~~F_{3}= p_{5}-p_{15},~~F_{4}= p_{8}-p_{5} \nonumber\\
F_{5}&=& p_{10}-p_{8},~~F_{6}= p_{13}-p_{10},~~F_{7}= p_{15}-p_{2}
\end{eqnarray}
The above information, along with the CS levels dictate that the baryonic symmetries are, in this case, given by the matrix
\begin{equation}
Q_D=
\begin{pmatrix}
p_{1} & p_{2} & p_{3} & p_{4} & p_{5} & p_{6} & p_{7} & p_{8} & p_{9} & p_{10} & p_{11} & p_{12} & p_{13} & p_{14} & p_{15} & p_{16} & p_{17} & p_{18}\\  
1 & 0 & 0 & 0 & 0 & 0 & 0 & 0 & 0 &-1 & 0 & 0 & 0 & 0 & 0 & 0 & 0 & 0\\
-1 & 1 & 0 & 0 & 0 & 0 & 0 & 0 & 0 & 0 & 0 & 0 &0 & 0 & 0 & 0 & 0 & 0\\ 
0 & 0 & 0 & 0 & 1 & 0 & 0 & 0 & 0 & 0 & 0 & 0 & 0 & 0 & -1 &0 & 0 & 0\\ 
0 & 0 & 0 & 0 & -1 & 0 & 0 & 1 & 0 & 0 & 0 & 0 & 0 & 0 & 0 & 0 & 0 & 0\\
1 & -1 & 0 & 0 & 0 & 0 & 0 & 0 & 0 & 0 & 0 & 0 & -1 & 0 & 1 & 0 & 0 & 0  
\end{pmatrix}
\end{equation}
We now obtain the data for the toric variety as
\begin{equation}
{\cal T}=
\begin{pmatrix}
0 & 0 & -1 & 1 & 0 & 0 & 0 & 0 & 0 & 0 & 0 & 0 & 0 & 0 & 0 & 1 & 0 & 0 \\ 
1 & 1 & 0 & 1 & 0 & -1 & 2 & 0 & 0 & 1 & 1 & -1 & 0 & 0 & 0 & 0  & 1 & 0\\
1 & 1 & 0 & 1 & 0 & 0 & 1 & 0 & 1 & 1 & 0 & -1 & 0 & 0 & 0 & 0 & 0 & -1
\end{pmatrix}
\end{equation}
This is seen to be toric variety which cannot be represented as a cone over a Fano 3-fold. The quiver diagram and the brane tiling picture for this theory is given 
in fig.(\ref{unhigf13}) and fig.(\ref{tilingf13}) respectively.

\begin{figure}[t!]
\begin{minipage}[b]{0.5\linewidth}
\centering
\includegraphics[width=2.8in,height=2.5in]{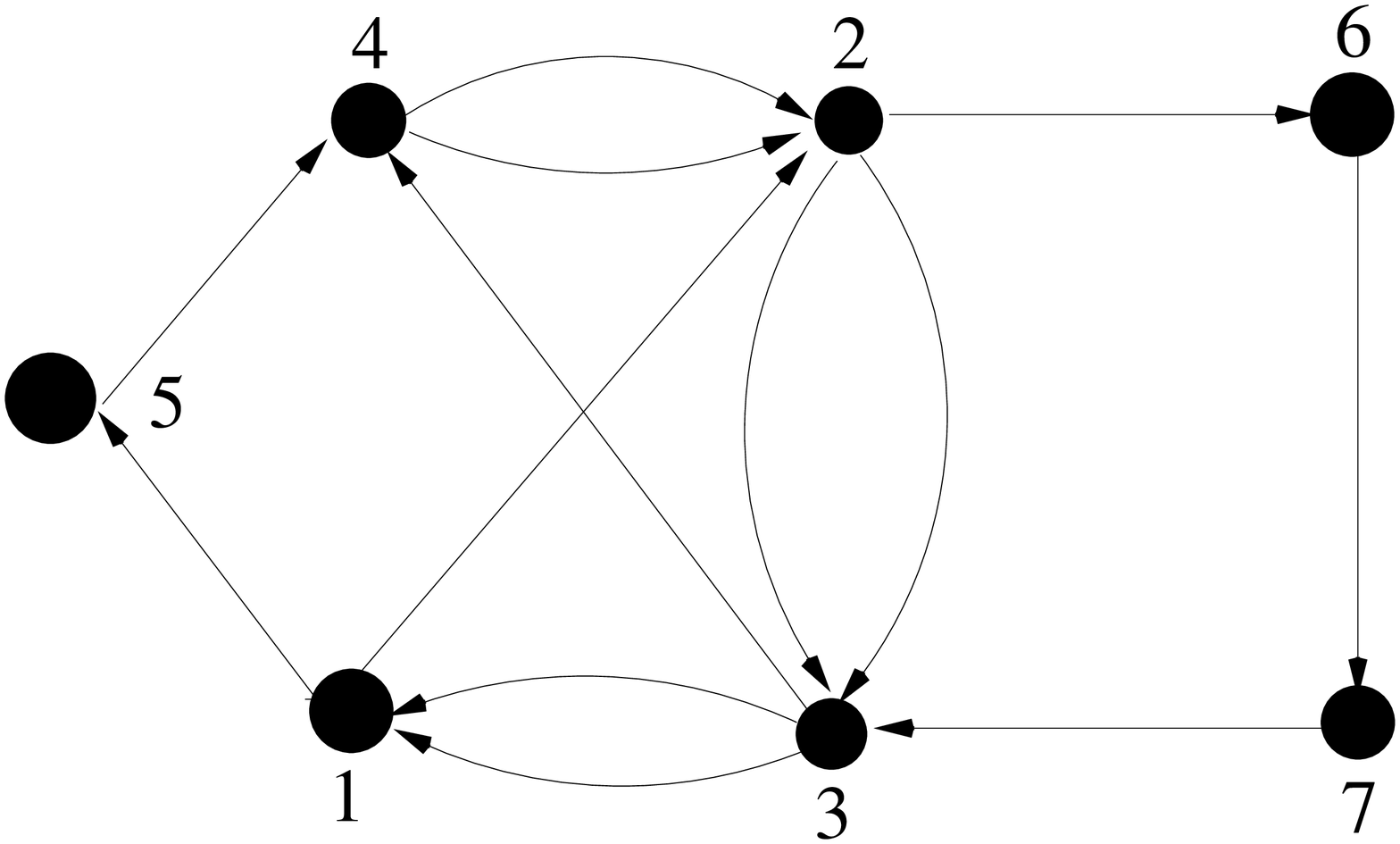}
\caption{Quiver diagram for the ${\mathcal F_2}$ theory unHiggsed by the field $X_{67}$. }
\label{unhigf21}
\end{minipage}
\hspace{0.6cm}
\begin{minipage}[b]{0.5\linewidth}
\centering
\includegraphics[width=3in,height=2.5in]{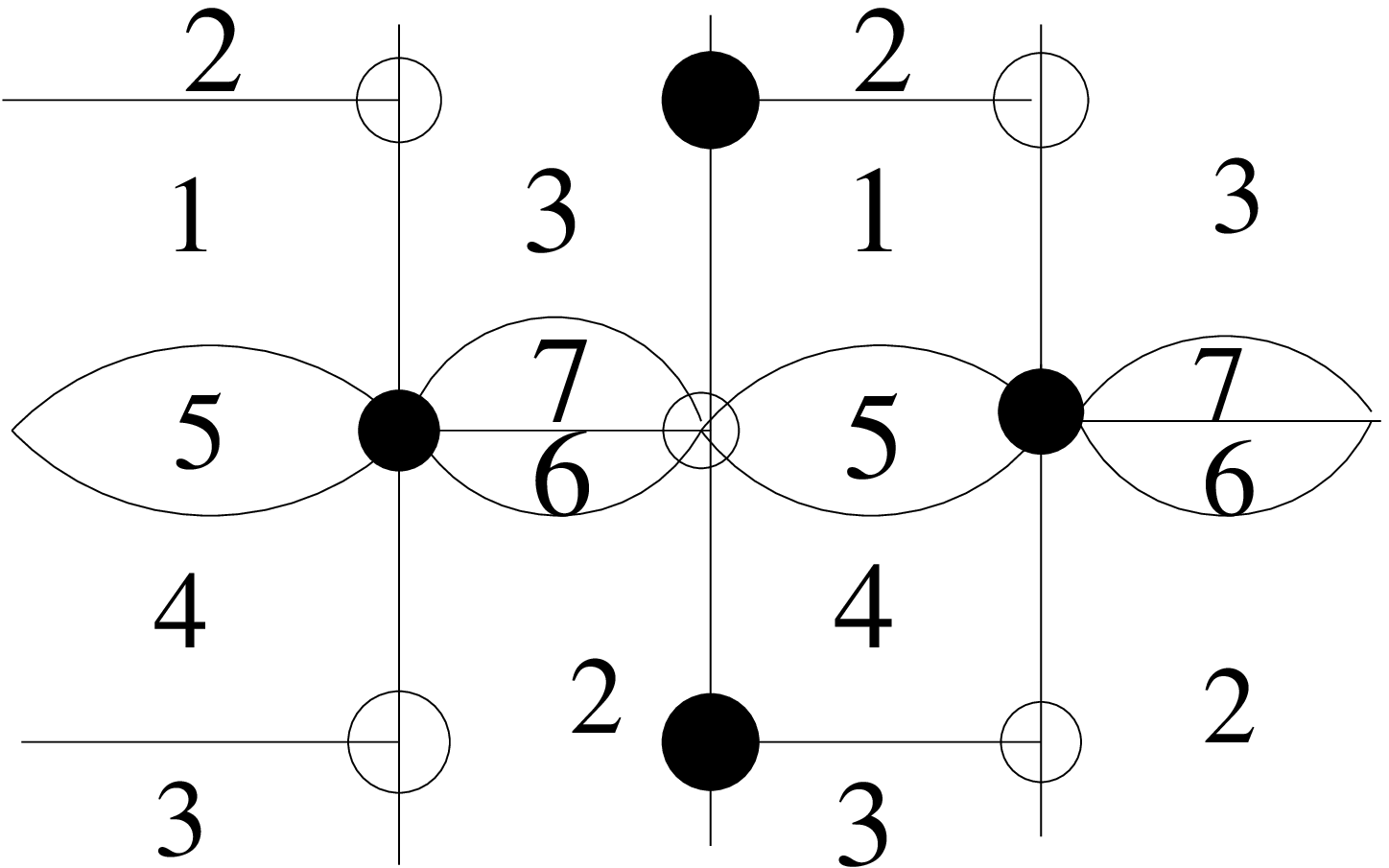}
\caption{Brane tiling for the ${\mathcal F_2}$ theory unHiggsed by the field $X_{67}$. }
\label{tilingf21}
\end{minipage}
\end{figure}

We now consider unHiggsing the theory ${\mathcal F_2}$. This model has 6 gauge groups and 12 chiral multiplates. The CS levels are, 
$\vec{k} = \left( 0, -1, 0, -1,1, 1\right)  $. The superpotential of the theory is given by,
\begin{eqnarray}
W &=& X_{12} X_{23}^{1} X_{31}^{2} + X_{34} X_{42}^{1} X_{23}^{2}  + X_{26}
X_{63} X_{31}^{1} X_{15} X_{54} X_{42}^{2}\nonumber\\
&-& X_{12} X_{23}^{2} X_{31}^{1} - X_{34}
X_{42}^{2} X_{23}^{1}  - X_{26} X_{63} X_{31}^{2} X_{15} X_{54} X_{42}^{1}
\end{eqnarray}
The toric data can be obtained from the quiver and the superpotential, and is given as
\begin{equation}
{\cal T}=
\begin{pmatrix}
1 & -1 & 0 & 0 & 0 & 0 & 0 & 0 & 0 & 0 & 0 & 0 \\ 
0 & 1 & 1 & -1 & 1 & -1 & 0 & 0 & 0 & 0 & 0 & 0 \\  
0 & 0 & 0 & 0 & 1 & -1 & -1 & 1 & 0 & 0 & 0 & 0
\end{pmatrix}
\label{f2toric1}
\end{equation}
As before, let us effect an unHiggsing by one field, which we choose to be  $X_{67}$. The resulting theory will have 7 gauge groups and 13 bifundamentals.
Choosing the CS levels to be $\vec{k} = \left( 0, -1, 0, -1,1, 1,0 \right)$, the superpotential is modified as 
\begin{eqnarray}
W &=& X_{12} X_{23}^{1} X_{31}^{2} + X_{34} X_{42}^{1} X_{23}^{2}  + X_{26}X_{67} X_{73} X_{31}^{1} X_{15} X_{54} X_{42}^{2}\nonumber\\
&-&X_{12} X_{23}^{2} X_{31}^{1} -X_{34} X_{42}^{2} X_{23}^{1}  - X_{26} X_{67} X_{73} X_{31}^{2} X_{15} X_{54}X_{42}^{1}
\end{eqnarray}
The final toric data can be represented in the form 
\begin{equation}
{\cal T}=
\begin{pmatrix}
0 & 0 & 0 & 0 & 0 & -1 & 1 & 0 & 0 & 0 & 0 & 0 & 0 & 0  \\  
1 & 0 & 0 & -1 & 0 & 1 & 0 & 0 & 1 & 0 & 0 & -1 & 0 & -1 \\  
0 & -1 & 0 & -1 & 0 & 0 & 0 & 0 & 1 & 1 & 0 & 0 & 0 & 0
\end{pmatrix}
\label{f2toric2}
\end{equation}
Comparing eq. (\ref{f2toric1}) with eq. (\ref{f2toric2}), we see that these are equivalent, apart from different multiplicities. 
The quiver diagram and the brane tiling picture of the unHiggsed theory is presented in fig.(\ref{unhigf21}) and fig.(\ref{tilingf21}) respectively.

Now, if we choose the CS levels as
$\vec{k} = \left( 0, -1, 0, -1, 1, 2,-1 \right)$ (which is also a consistent choice of CS levels, given those of the original theory), the situation changes. In this case, we find that 
\begin{equation}
Q_D =
\begin{pmatrix}
p_{1} & p_{2} & p_{3}
& p_{4} & p_{5} & p_{6} & p_{7} & p_{8} & p_{9} & p_{10} & p_{11} & p_{12} & p_{13}& p_{14}  \\  
0 & 0 & 1 & 0 & -1 & 0 & 0 & 0 & 0 & 0 & 0 & 0 & 0 & 0 \\
0 & 0 & 0 & 0 & 1 & 0 & 0 & 0 & 0 & 0 & 0 & 0 & -1 & 0 \\ 
0 & 0 & -1 & 0 & 0 & 0 & 0 & 1 & 0 & 0 & 0 & 0 & 0 & 0 \\  
0 & 1 & -1 & 0 & 0 & 0 & 0 & -1 & 0 & 1 & 0 & 0 & 0 & 0\\ 
0 & 0 & -1 & 0 & 0 & 0 & 0 & 0 & 0 & 1 & -1 & 0 & 1 & 0
\end{pmatrix}
\end{equation}
which can now be seen to give the toric four-fold with data
\begin{equation}
{\cal T}=
\begin{pmatrix}
0 & 0 & 0 & 0 & 0 & -1 & 1 & 0 & 0 & 0 & 0 & 0 & 0 & 0\\
1 & 0 & 0 & -1 & 0 & 1 & 0 & 0 & 1 & 0 & 0 & -1& 0 & -1\\ 
0 & -1 & 0 & -1 & 0 & 0 & 0 & 0 & 1 & 1 & 1 & 1 & 0 & 0
\end{pmatrix}
\end{equation}
As in our previous example, this is a toric four-fold which is not a cone over a Fano threefold.  

As a further example, we consider unHiggsing the theory ${\mathcal E_4}$. We will be brief here. The original theory is one of 5 gauge groups and 9 chiral multiplets
with the CS levels chosen to be $\vec{k} = \left( 1, -1, 0,-1, 1\right)$. Upon unHiggsing by the field $X_{46}$, the superpotential of the theory is modified to
\begin{eqnarray}
W &=& X_{12}^{1} X_{23}^{1} X_{34}^{2} X_{45} X_{51} +X_{12}^{2} X_{23}^{2}X_{34}^{1} X_{46} X_{61} \nonumber\\
&-& X_{12}^{2} X_{23}^{1} X_{34}^{1} X_{45} X_{51}-X_{12}^{1}X_{23}^{2} X_{34}^{2} X_{46} X_{61}
\end{eqnarray}
The matching matrix contains 13 perfect matchings with each involving 2 bifundamental fields, and one calculates the baryonic symmetries from the 
face symmetries and the CS levels, which we choose to be $\vec{k} = \left( 1, -1, 0, 0, 1,-1 \right)$. Concatenating with the matrix $Q_F$ obtained as the kernel of the 
perfect matching matrix, we finally obtain the toric data for this variety as 
\begin{equation}
{\cal T}=
\begin{pmatrix}
0 & -1 & 1 & 0 & 0 & 0 & 0 & 0 & 0 & 0 & 0 & 0 & 0\\
1 & 1 & 0 & 0 & 1 & 0 & -1 & 0 & 0 & 0 & -1 & 1 & 0\\ 
0 & 0 & 0 & 0 & -1 & -1 & 1 & 1 & 0 & 0 & 0 & 0 & 0 \\ 
\end{pmatrix}
\end{equation}
This is again an example of a toric four-fold that is not Fano, obtained by unHiggsing a Fano 3-fold. 

Finally, we study unHiggsing of toric four-folds with adjoint fields that result in complex cones over Fano 3-folds. Consider the theory with 
3 gauge groups, 6 bi-fundamental and 1 adjoint field. We have labeled the gauge groups as 1, 3 and 4 for later conveniance. The CS levels are
chosen to be $\vec{k} = \left( 0, -1, 1\right)$, and the superpotential is taken to be 
\begin{equation}
W= X_{11} X_{13}^{1} X_{34}^{1} X_{41}^{2} +X_{13}^{2} X_{34}^{2}
X_{41}^{1} -X_{11} X_{13}^{2} X_{34}^{1} X_{41}^{1} -X_{13}^{1} X_{34}^{2}
X_{41}^{2}
\end{equation}
where $X_{11}$ is the adjoint. A simple calculation shows that this is a toric four-fold, with toric data 
\begin{equation}
{\cal T}=
\begin{pmatrix}
0 & 0 & 0 & 1 & 0 & 0 \\ 
0 & -1 & 1 & 0 & 0 & 0 \\ 
1 & 0 & 0 & 0 & 0 & 0
\end{pmatrix}
\end{equation}
We now study the unHiggsing of this theory by introducing an extra gauge group, which we label as 2, with an additional field $X_{12}$. 
The theory obtained will have 4 gauge groups and 8 fields. Other than the added field, the adjoint field over the gauge group 1 gets transformed into $X_{12}$ and 
the $X_{13}^{i}$s transform to $X_{23}^{i}$s. We choose the CS levels to be, $\vec{k} = \left( 1 , -1 ,-1, 1 \right)$. 
The superpotential of the new theory is taken as
\begin{equation}
W =X_{12}^{2} X_{23}^{1} X_{34}^{1} X_{41}^{2} + X_{12}^{1} X_{23}^{2}
X_{34}^{2} X_{41}^{1} -X_{12}^{2} X_{23}^{2} X_{34}^{1} X_{41}^{1} - X_{12}^{1}
X_{23}^{1} X_{34}^{2} X_{41}^{2}
\end{equation}
From the superpotential, the perfect matching matrix is calculated as 
\begin{equation}
P=
\begin{pmatrix}
& p_{1} & p_{2} & p_{3} & p_{4} & p_{5} & p_{6} & p_{7} & p_{8} \\ 
X_{12}^{1} & 1 & 0 & 1 & 0 & 0 & 0 & 0 & 0 \\ 
X_{12}^{2} & 1 & 0 & 0 & 1 & 0 & 0 & 0 & 0 \\
X_{23}^{1} & 0 & 1 & 0 & 0 & 0 & 1 & 0 & 0 \\
X_{23}^{2} & 0 & 1 & 0 & 0 & 0 & 0 & 1 & 0 \\ 
X_{34}^{1} & 0 & 0 & 1 & 0 & 1 & 0 & 0 & 0\\ 
X_{34}^{2} & 0 & 0 & 0 & 1 & 1 & 0 & 0 & 0 \\ 
X_{41}^{1} & 0 & 0 & 0 & 0 & 0 & 1 & 0 & 1 \\ 
X_{41}^{2} & 0 & 0 & 0 & 0 & 0 & 0 & 1 & 1
\end{pmatrix}
\end{equation}
From the tiling picture, we can calculate the face symmetries
\begin{equation}
F_{1}= p_{1}-p_{8},~~F_{2} = p_{2}-p_{1},~~F_{3}= p_{5}-p_{2},~~F_{4}= p_{8}-p_{5} 
\end{equation}
The matrix $Q_F$ is obtained as the kernel of the perfect matching matrix, whereas the baryonic symmetries are given by the matrix
\begin{equation}
Q_D=
\begin{pmatrix}
 p_{1} & p_{2} & p_{3} & p_{4} & p_{5} & p_{6} & p_{7} & p_{8} \\ 
 0 & 1 & 0 & 0 & 0 & 0 & 0 & -1\\ 
 1 & -1 & 0 & 0 & 1 & 0 & 0 & -1
\end{pmatrix}
\end{equation}
and the toric data can be cast in the form 
\begin{equation}
{\cal T}=
\begin{pmatrix}
0 & 0 & 0 & 0 & 0 & -1 & 1 & 0 \\ 
-1 & 0 & 0 & 0 & 1 & 0 & 0 & 0 \\ 
0 & 0 & -1 & 1 & 0 & 0 & 0 & 0 \\
\end{pmatrix}
\end{equation}
This is seen to be phase I of the variety ${\mathcal C_3}$. 

Finally, let us consider unHiggsing with an adjoint field for the theory of 3 gauge groups (which we call 1, 3 and 4), 6 bi-fundamental and 1 adjoint field, where we take the 
CS levels to be $\vec{k} = \left( -1, 1, 0\right)$. The superpotential is chosen as 
\begin{equation}
W= X_{11} X_{13}^{2} X_{34}^{1} X_{41}^{1} +X_{13}^{1} X_{34}^{2}
X_{41}^{2} -X_{11} X_{13}^{1} X_{34}^{1} X_{41}^{2} -X_{13}^{2} X_{34}^{2}
X_{41}^{1}
\end{equation}
It can be verified that this theory, upon unHiggsing by adding a field $X_{12}$ (i.e adding a gauge group which we label as 2) gives rise to phase I of the
theory ${\mathcal C}_5$. 

Let us now summarise the results in this section. We  have first studied the unHiggsing of the theory ${\mathcal F}_1$. We have shown that taking a particular
choice of CS levels consistent with the original theory, unHiggsing by one field might results in the same toric variety probed by the M2-brane theory, although with
multiplicities different from the parent theory. This is similar to the case of the Fano 2-folds studied in section 2, and here also we see that the corner points of the 
toric diagrams acquire multiplicities. A different choice of CS levels gives a non-Fano theory which is a CY 4-fold that has not previously appeared in the literature,
where in some of the cases the singularity seems to be non-isolated. 
We also performed an explicit computation of unHiggsing the ${\mathcal F}_1$ theory with three fields, and found a new non-Fano CY 4-fold. Similar analyses were
performed for the varieties ${\mathcal F}_2$ and ${\mathcal E}_4$. We also performed unHiggsing of theories with adjoints, extending the results of \cite{bhs} to reach 
certain phases of the known theories ${\mathcal C}_3$ and ${\mathcal C}_5$. These are the main results of this section. 

\section{M2-brane Theories Without a Brane Tiling Description}

In this final section of the paper, we will study an interesting example of an M2-brane theory that does not seem to admit a brane tiling description. The origin of
the results presented below is a study of the inverse algorithm, which in its original {\it avatar} appeared in the work \cite{fhh}, and attempts to construct a sensible gauge
theory living on a brane from the toric data of the singularity which it probes. Since this algorithm, for M2-brane theories is far
from clear (for a recent discussion on the algorithm, see \cite{rama}), we will postpone a discussion on this for the conclusion. We simply begin with 
the theory ${\mathcal B_1}$ \cite{fanohan} with added multiplicities (in hindsight), such that the toric data is given by
\begin{equation}
{\cal T}=
\begin{pmatrix}
1 & -1 & 0 & 0 & 0 & 0 & 0 & 0 \\ 
0 & 1 & -1 & 0 & 0 & 0 & 0 & 0 \\ 
0 & 0 & 2 & -1 & 1 & 0 & 1 & 0 \\
\end{pmatrix}
\end{equation}
The total charge matrix is obtained as the kernel of the toric data, and is given as 
\begin{equation}
Q =
\begin{pmatrix}
1 & 1 & 1 & 0 & -1 & 0 & -1 & -1 \\ 
0 & 0 & 0 & 0 & 0 & 1 & 0 & -1 \\ 
0 & 0 & 0 & 1 & 0 & -1 & 1 & -1 \\
0 & 0 & 0 & 0 & 1 & 0 & -1 & 0
\end{pmatrix}
\label{notilinga}
\end{equation}
Let us assume that the first two rows denote the $Q_F$ and the others denote the $Q_D$  matrices respectively. The CS levels are taken to be
${\vec k} = \left(-1,2,0,-1\right)$. As far as the gauge theory is concerned, we have already made a departure from its description as a theory 
for ${\mathcal B}_1$, since the number of baryonic symmetries for such a theory should be one, as the second Betti number for the variety is 2. However, any choice
of a single column $Q_D$ in the above charge matrix does not seem to give a consistent $(2+1)$d theory. 
Note that in a sense,
we have drawn on the ``F-D ambiguity'' discussed in \cite{fhh}, although it is clear that the gauge theory will not have a tiling description. 
\begin{figure}[t!]
\begin{minipage}[b]{0.5\linewidth}
\centering
\includegraphics[width=2.8in,height=2.5in]{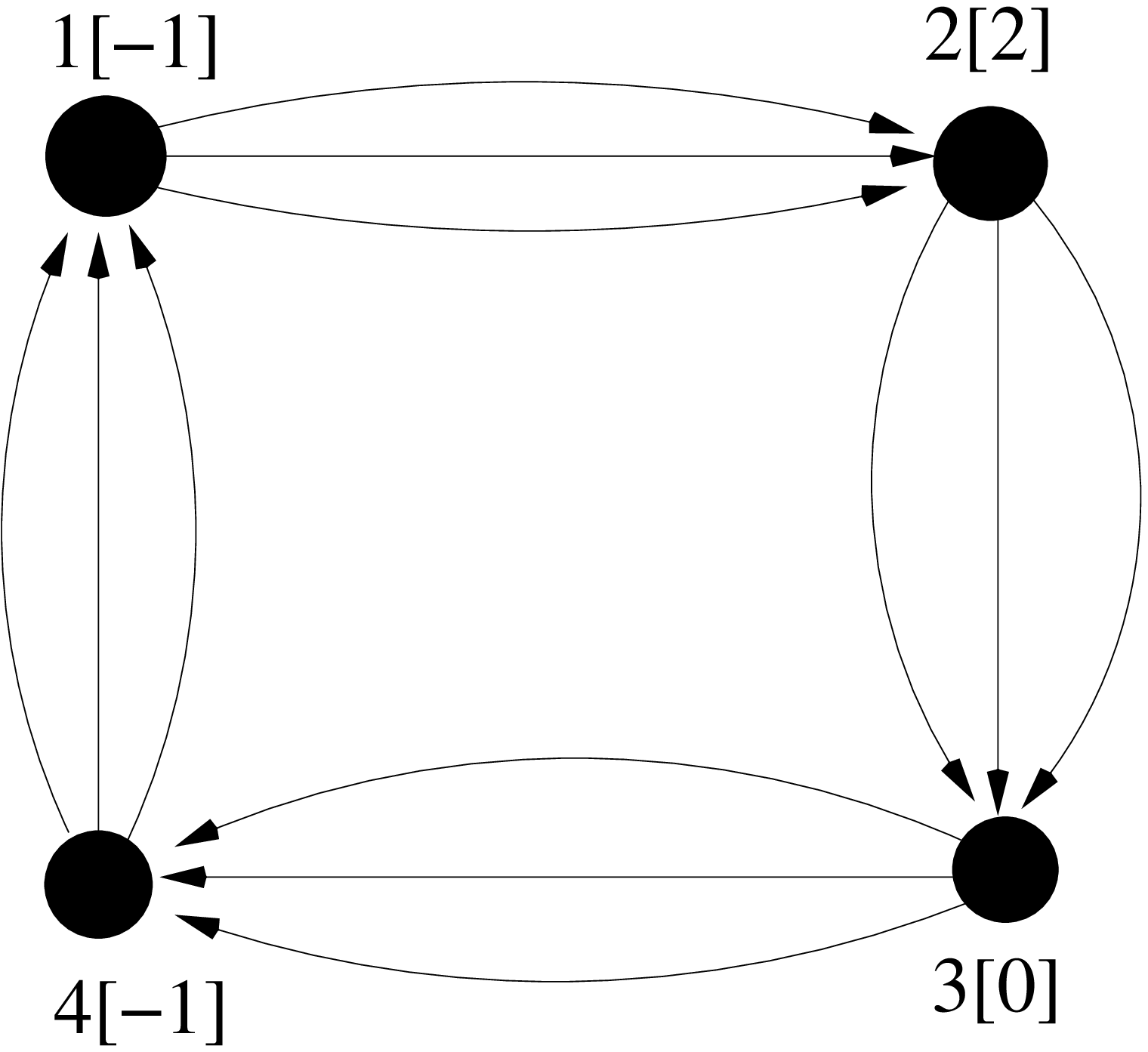}
\caption{Quiver diagram for the theory with charge matrix given by eq.(\ref{notilinga}). }
\label{notiling1}
\end{minipage}
\hspace{0.6cm}
\begin{minipage}[b]{0.5\linewidth}
\centering
\includegraphics[width=3in,height=2.5in]{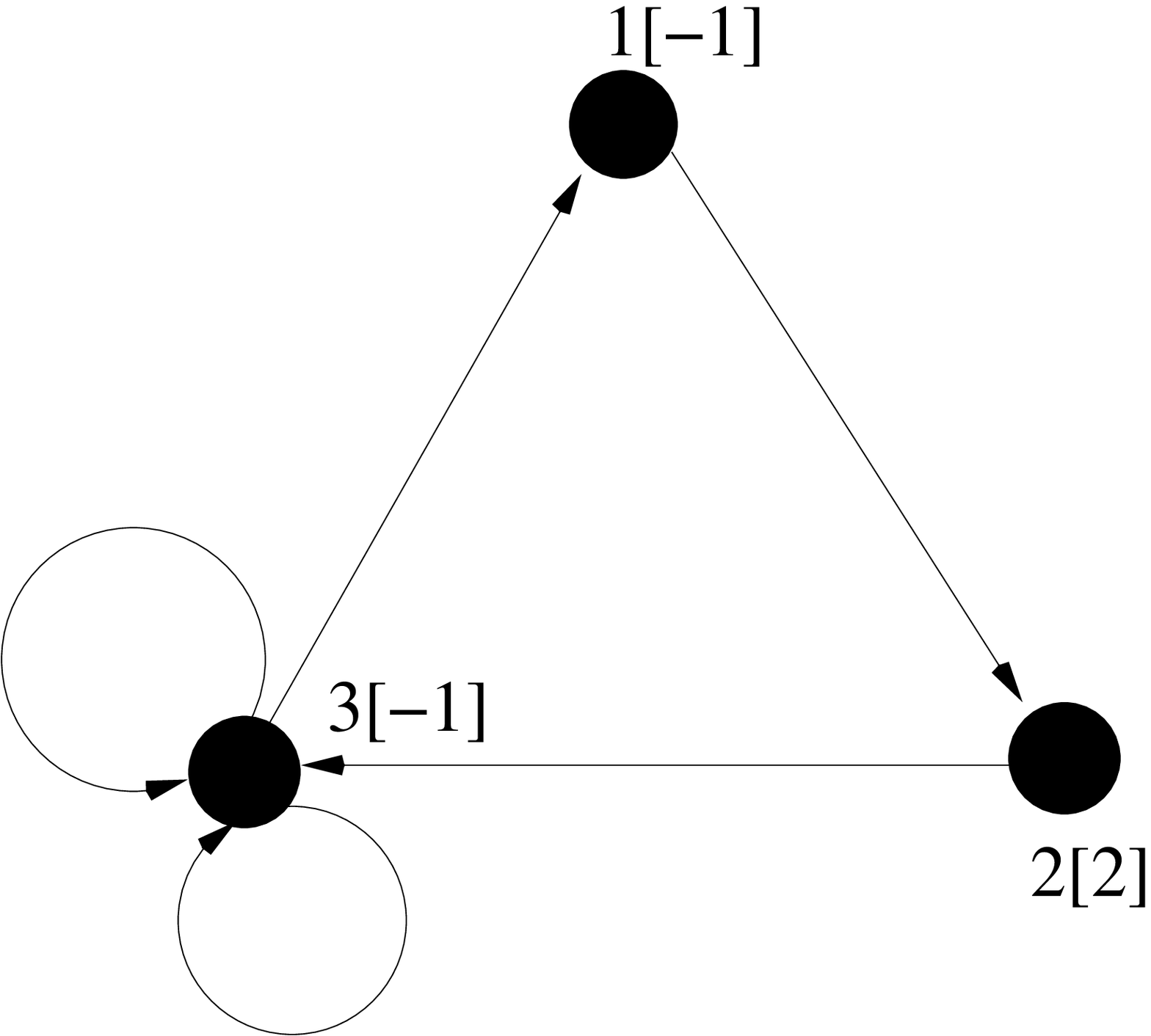}
\caption{Quiver diagram for the theory of fig.(\ref{notiling1}) Higgsed by the field $X_{34}^1$ }
\label{notiling2}
\end{minipage}
\end{figure}
The ``perfect matching matrix'' (by abuse of notation we will use the same terminology from the tiling picture, although the physical significance of the matrix $P$ is
unclear to us) from the given $Q_F$ can calculated. We mention here that in this case, two extra columns need to be added to the $K$ matrix to make it consistent.  
\footnote{The extra columns do not correspond to adjoint fields here, as seen from the quiver charge matrix.}
Whereas this might seem ad hoc, it allows for the construction of a superpotential and a consistent quiver diagram, with the given CS levels. 
The superpotential is obtained by integrating the $K$ matrix, and reads \footnote{We could have written down a superpotential for this theory that has a tiling description, but
this would result in a different CY 4-fold.}
\begin{eqnarray}
W &=& X_{12}^1\left(X_{23}^2X_{34}^1X_{41}^3 - X_{23}^1X_{34}^3X_{41}^2\right) + X_{12}^2\left(X_{23}^3X_{34}^2X_{41}^1 - X_{23}^2X_{34}^1X_{41}^3\right) \nonumber\\
&+& X_{12}^3\left(X_{23}^1X_{34}^3X_{41}^2 - X_{23}^3X_{34}^2X_{41}^1\right) 
\label{superpotnotile}
\end{eqnarray}
and the perfect matching matrix given by
\begin{equation}
P =
\begin{pmatrix}
0&0&0&1&0&0&0&0\\
1&0&0&0&1&0&0&0\\
0&1&0&0&1&0&0&0\\
0&0&1&0&1&0&0&0\\
1&0&0&0&0&0&1&0\\
0&1&0&0&0&0&1&0\\
0&0&1&0&0&0&1&0\\
1&0&0&0&0&1&0&1\\
0&1&0&0&0&1&0&1\\
0&0&1&0&0&1&0&1\\
0&0&0&1&0&0&0&0\\
0&0&0&1&0&0&0&0
\end{pmatrix}
\end{equation}
where the bifundamental indices in eq.(\ref{superpotnotile}) is incorporated from the quiver charge matrix which can be shown to be given by
\begin{equation}
d=
\begin{pmatrix}
1&0&0&0&0&0&0&-1&-1&-1&1&1\\
-1&0&0&0&1&1&1&0&0&0&-1&-1\\
0&1&1&1&-1&-1&-1&0&0&0&0&0\\
0&-1&-1&-1&0&0&0&1&1&1&0&0
\end{pmatrix}
\label{quivernotile}
\end{equation}
The quiver is shown in fig.(\ref{notiling1}). In this theory, the usual tiling condition $E = G + N_T$ is not satisfied and it is not possible to construct a tiling from
eq.(\ref{superpotnotile}). But note that if we regard the matrix ${\cal P}$ as given, then in conjunction with the quiver data of eq.(\ref{quivernotile}), we can construct the
toric data, eq.(\ref{notilinga}) of the variety ${\mathcal B}_1$. We therefore interpret this as a CY 4-fold theory that does not have a tiling picture. 

\begin{figure}[t!]
\centering
\includegraphics[width=3in,height=2.5in]{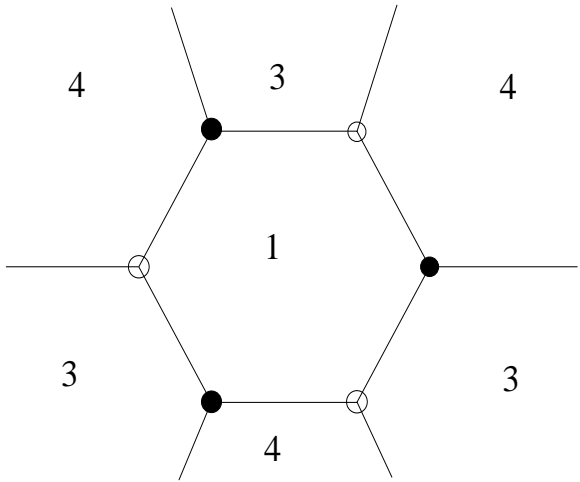}
\caption{Brane tiling for Higgsing the theory described by eq.(\ref{superpotnotile}).}
\label{notilingfig}
\end{figure}
Now let us consider Higgsing the theory with vevs to the three fields $X_{12}^i$. \footnote{Note that here the fields $X_{12}^i$ appear with the same perfect matching. Higgsing
one of them results in an inconsistent theory.} From the superpotential, we see that the quiver reduces to that of the theory 
${\mathcal B_4}$ (also called $M^{1,1,1}$) \cite{fanohan}, but now with CS levels ${\vec k} = \left(1,0,-1\right)$. These values of the CS levels mean that the theory is not 
the ${\mathcal B_4}$ theory of \cite{hanhigg}, although it has the same quiver and tiling description. The brane tiling for this case (which is the same as that of the theory 
${\mathcal B}_4$) is shown in fig.(\ref{notilingfig}). The resulting theory is a CY 4-fold which is not Fano, and has toric data 
\begin{equation}
{\cal T}=
\begin{pmatrix}
0 & 1 & -1 & 0 & 0 & 0 & 0  \\ 
-1 & 0 & 1 & 0 & 0 & 0 & 0  \\ 
2 & 0 & 0 & 1 & 0 & 1 & 0  
\end{pmatrix}
\end{equation} 

We also consider Higgsing the original theory by removing the field $X_{34}^1$. The resulting theory
can be shown to be of a CY 4-fold which has a vanishing superpotential and the quiver diagram is presented in fig.(\ref{notiling2}). This toric variety corresponds to the data
\begin{equation}
{\cal T}=
\begin{pmatrix}
1&-1&0&0&0\\
0&1&0&0&0\\
0&0&-1&1&0
\end{pmatrix}
\end{equation} 

To summarise, in this section, we have considered an example of an M2-brane theory which may not be represented by a tiling description. We have used an analogue
of the ``F-D'' ambiguity in the inverse algorithm of \cite{fhh} to arrive at the theory. Specifying the quiver diagram and the superpotential, we have seen that the latter cannot
give rise to a consistent brane tiling picture. To analyse the theory, we have used the ``perfect matching matrix'' which is obtained by the $K$ and the $T$
matrices that arise from choosing the F- and D-term symmetries from the total charge matrix. Higgsing the theory, we found a resulting CY 4-fold which does admit of
a tiling description. These are the main results of this section. Note that theories that do not admit of a brane tiling picture have not been dealt with in the literature so far in 
a systematic way. Admittedly, our analysis here is preliminary, and far from being complete. The issue needs to be studied in much more details, as we discuss in the final section. 

\section{Conclusions}

In this paper, we have addressed the issue of Higgsing and unHiggsing of certain $(2+1)$d CS theories that are conjectured to be theories of M2-branes probing cones over smooth
toric Fano 3-folds. The main results presented in this paper are summarised as follows. In section 2, we have shown that unHiggsing known Fano 2-folds by one field sometimes
give rise to the same toric diagram, but with corner point multiplicities. This implies that the
corresponding gauge theories cannot be obtained by partial resolutions of the theory $\BC^3/\BZ_3\times\BZ_3$, and 
constitute new examples of D3-brane world-volume theories. In section 3, we studied in details the Higgsing procedure for some of the smooth toric Fano 3-folds. We have established
a few of the blowup-blowdown relations that exist between these varieties, although our results suggest that there are possibly more phases of the theories presented in
\cite{fanohan} to reproduce all the known connections between these. We have also seen that in some cases, Higgsing of a Fano 3-fold results in a theory that is not Fano. 
An interesting example was the emergence of the theory $\BP^1\times {\rm SPP}$ from Higgsing the theory ${\mathcal F}_1$ by a single field. 
In section 4, we have studied unHiggsing of some Fano 3-folds, and as in the case of the del Pezzo surfaces, we see that there are theories with different matter content 
and superpotential that describe the same Fano variety, but with different multiplicities in the toric diagram. These are new phases of the brane theories. We have also 
discussed some examples of non-Fano theories (for possibly non-isolated singularities) that arise from the Fano 3-folds for different choices of the CS levels. 
Further, we have performed unHiggsing of $(2+1)$d theories with adjoint fields 
to reach known phases of the theories ${\mathcal C}_3$ and ${\mathcal C}_5$. Finally, in section 5, we have attempted a construction of a M2-brane gauge
theory that does not seem to have a tiling description, and examined the Higgsing of this theory. We believe that these results constitute the first steps in understanding the
complex web of relationships that are known to exist in the mathematics literature, between smooth Fano 3-folds. 

We end this paper with a discussion of the ``inverse algorithm'' for M2-brane theories. In its original form, the inverse algorithm of \cite{fhh} (for gauge theories on
D3-branes) attempts to embed a given toric diagram (for which one wants to construct the corresponding gauge theory) into a known singularity and removes points from the
toric diagram (i.e perfect matchings), to reach the singularity in question.  In the process, the quiver charge matrix and the superpotential of the theory in question can be 
determined. The embeddings alluded to here can always be performed for D3-brane theories. For M2-brane theories, such an inverse algorithm has not yet been 
studied systematically. One can, however, start with the toric data for a given theory and proceed by constructing a charge matrix from its kernel. Then one has to decide which
of the rows of the resulting charge matrix denote the F-terms and which denote the D-terms. Clearly, there is a huge ambiguity in the process, with an additional complication
being present due to possible multiplicities in the toric diagram. For most of the 14 Fano 3-fold theories presented in \cite{fanohan}, we have checked that one can, in fact,
construct the M2-brane gauge theory for one unique choice of the $Q_D$ and $Q_F$ matrices (where the number of baryonic symmetries is determined from the second Betti number
of the variety), but this fails precisely for the theories ${\mathcal B}_{1,2,3}$ where no 
consistent gauge theory could be constructed in this method. The results of the last section of this paper shows that in this context, it might be interesting to look at 
theories that might not admit of a tiling description. 

It will be interesting to understand the full class of string theoretic relations between the smooth Fano 3-folds, as well as the inverse algorithm for M2-brane theories. These
are important directions for research on the subject, and we hope to address these issues in a future publication. 

\begin{center}
{\bf Acknowledgements}
\end{center}
We gratefully acknowledge email correspondence with S. Dwivedi and P. Ramadevi on issues related to the ones discussed in this paper.

\end{document}